\documentclass[aps,prb,preprint,groupedaddress,superscriptaddress,showpacs,floatfix]{revtex4-1}
\usepackage{mhchem}
\usepackage{textcomp}
\usepackage{multirow}
\usepackage{amsmath,amsfonts,amssymb}
\usepackage{graphicx}
\usepackage{dcolumn}   
\usepackage{bm}        
\usepackage{longtable}
\usepackage{tabularx}
\usepackage{color}
\begin{document}

\title{Hybrid Density Functional Study of Structural and Electronic Properties of Functionalized
\ce{Ti_{n+1}X_n} (X= C, N) monolayers}
\author{Yu Xie}
\email{yxe@ornl.gov}
\affiliation{Center for Nanophase Materials Sciences, Oak Ridge National 
Laboratory, Oak Ridge, Tennessee 37831, USA}
\author{P. R. C. Kent}
\affiliation{Center for Nanophase Materials Sciences, Oak Ridge National 
Laboratory, Oak Ridge, Tennessee 37831, USA}
\affiliation{Computer Science and Mathematics Division, Oak Ridge National 
Laboratory, Oak Ridge, Tennessee 37831, USA}
\date{\today}

\begin{abstract}
Density functional theory simulations with conventional (PBE) and 
hybrid (HSE06) functionals
were performed to investigate the structural and electronic properties of
MXene monolayers, \ce{Ti_{n+1}C_n} and \ce{Ti_{n+1}N_n}
($n$ = 1--9) with surfaces terminated by O, F, H, and OH groups.
We find that PBE and HSE06 give similar results. Without
functional groups, MXenes have magnetically ordered ground states.
All the studied materials are metallic except for \ce{Ti_{2}CO_{2}},
which we predict to be semiconducting.
The calculated density of states at the Fermi level of the thicker MXenes
($n$ $\geqslant$ 5) is much higher than for thin MXenes,  indicating
that properties such as electronic conductivity and surface chemistry will be
different. In general, the carbides and nitrides behave
differently with the same functional groups.
\end{abstract}

\maketitle

\section{\label{sec:intro}Introduction}
The study of two-dimensional (2D) materials is a topic of current interest,
not only for their unique properties\cite{novoselov2004electric,berger2006electronic,
zhang2005experimental} compared with their three-dimensional
(3D) counterparts, but also for their important applications to industry 
and engineering\cite{stoller2008graphene,yoon2011good,neto2009electronic,
yoo2011ultrathin}. Over the years, despite many experimental efforts on
2D materials, only a few classes of freestanding single-layer 2D materials have been
successfully synthesized. Among them, the very first and most fascinating
one is graphene\cite{novoselov2004electric} which is prepared from 
exfoliated graphite. Other graphene-like
 materials, such as hexagonal BN and dichalcogenides, 
are synthesized by exfoliation of their 3D-layered precursors.\cite{ramakrishna2010mos2,
pacile2008two,novoselov2005two} Normally, these compounds are considered as van der
Waals solids due to the weak interaction between layers, making
them able to be exfoliated. Alternatively, the synthesis of freestanding
2D monolayers from 3D-layered solids with strong interlayer bonds
is more difficult.

\begin{figure}
\centering
\includegraphics[scale=1.5]{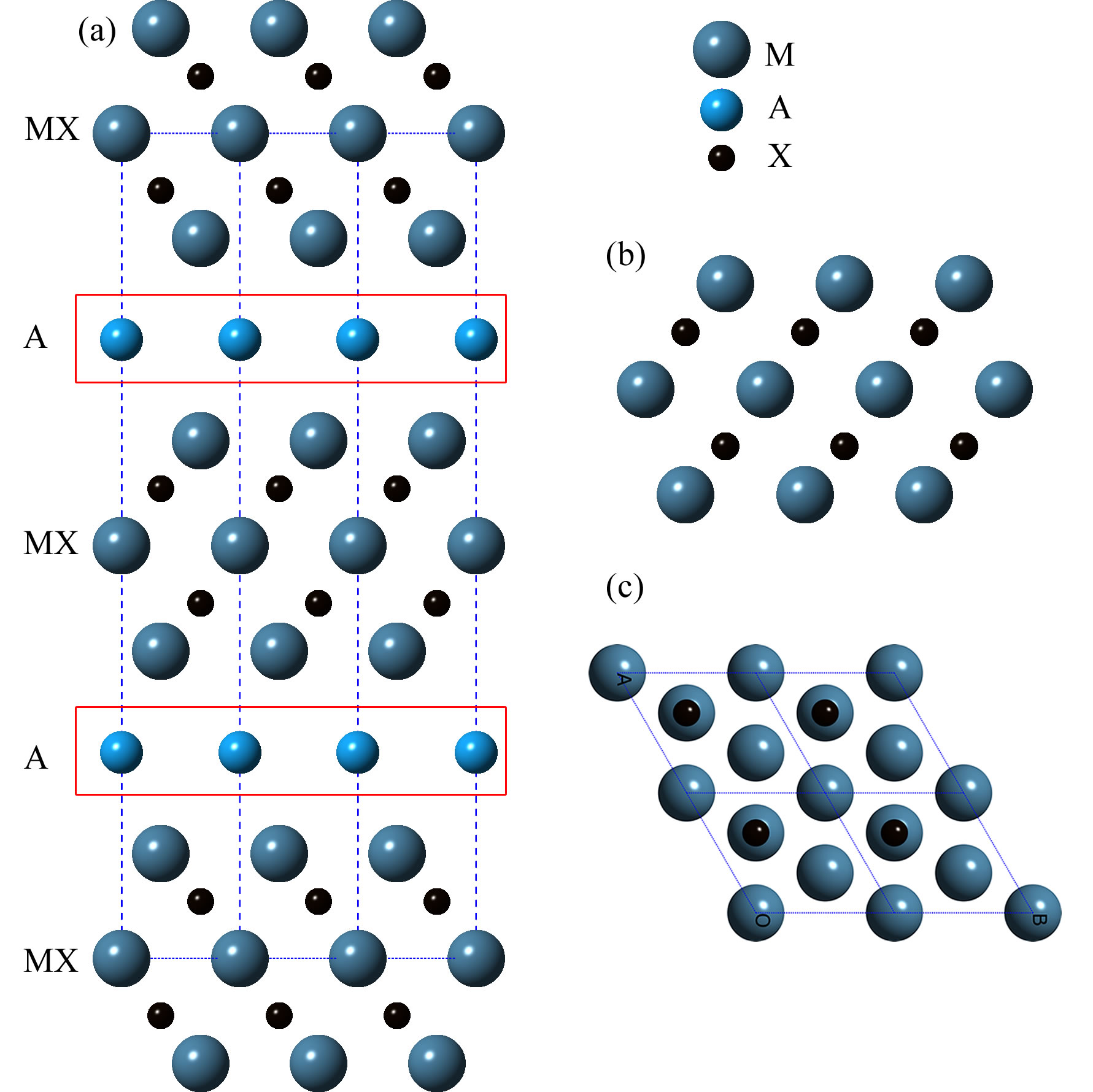}
\caption{(color online) (a) Crystal structure of  layered \ce{M_{3}AX_{2}} solid phase.
The  \ce{M_{3}X_{2}} and A layer are stacked in a zigzag formation.
(b) and (c) Side and top views of \ce{M_{3}X_{2}} monolayer.}
\label{fig:MAX}
\end{figure}

Very recently, a new family of 2D materials was prepared by the 
exfoliation of the layered ternary transition metal carbides,
which are known as MAX phase\cite{eklund2010m,barsoum2000mn+}. 
The latter is a large family
of layered solids, including ternary transition metal
carbides, nitrides, and carbonitrides, with more than 70
members. The MAX phases usually possess unique properties, such as
remarkable machinability, high damage tolerance, excellent oxidation
resistance, and high electrical and thermal conductivity, with
various industrial applications.\cite{barsoum2011elastic,wang2010layered,sun2011progress} 
These solids have a general formula of  \ce{M_{n+1}AX_{n}}
($n$ = 1, 2, 3, $\cdots$), where ``M" is an early transition metal,
``A" represents the main group elements (mostly group IIIA
and IVA), and ``X" denotes C and/or N.  MAX phases adopt
hexagonal structures (space group $P$63/$mmc$), and their structures
can be viewed as inter-growth structures consisting
of hexagonal \ce{M_{n+1}X_{n}} layers and planar A 
atomic sheets with alternative stacking along the $c$ direction.
The crystal structure of  \ce{M_{3}AX_{2}} is shown in Fig. \ref{fig:MAX}(a) as 
an example. In general, the MAX phases are quite stable.
However, compared to the strong M--X intralayer bonds, the 
interatomic A--A bonds and interlayer M--A bonds are 
weaker, making A atomic layers chemically more reactive. In turn,
taking \ce{Ti_{3}AlC_{2}} as an example,
the Al layers can be extracted from the solid \ce{Ti_{3}AlC_{2}} 
by hydrofluoric acid treatment and sonication,
resulting in graphene-like \ce{Ti_{3}C_{2}}
nanosheets, so-called ``MXene".\cite{naguib2011two} 
Depending on the synthesis conditions,
the surfaces of \ce{Ti_{3}C_{2}} monolayers are chemically terminated 
with oxygen-containing
and/or fluoride groups. Following the same procedure, 
other MXenes, including \ce{Ti_{2}C}, 
\ce{Ta_{4}C_{3}}, \ce{Ti_{3}CN}, 
(\ce{Ti_{0.5}Nb_{0.5})_{2}C},
and (\ce{Ti_{0.5}Cr_{0.5})_{2}C},
have been prepared by exfoliation of the corresponding carbides
and carbonitrides.\cite{naguib2012two12} 
Notably, due to the large number of MAX family, it is predictable that more
MXenes will be synthesized  by exfoliation of their
corresponding MAX phases.

Several studies\cite{Enyashin201227,kurtoglu2012first,come2012non,naguib2012mxene,
tang2012mxenes,mashtalir2013inter,khazaei2012novel,naguib2011two,naguib2012two12,
shein2012graphene,lane2013correlation} 
have been conducted to understand the properties of
these newly discovered graphene-like MXenes. It has been shown that
these materials are good electrical conductors\cite{naguib2012two12,naguib2011two,Enyashin201227} 
and have high
elastic moduli\cite{kurtoglu2012first}. The intercalation of Li ion 
and organic molecules into 
\ce{Ti_{2}C}\cite{come2012non,naguib2012mxene} 
, \ce{Ti_{3}C_{2}}\cite{tang2012mxenes,mashtalir2013inter},  
\ce{Ti_{3}CN}\cite{mashtalir2013inter}, and \ce{TiNbC}\cite{mashtalir2013inter}
sheets suggests they are promising
candidates for Li-ion battery anodes and hybrid electrochemical
capacitors. And with proper termination groups, the functionalized 
MXenes could become semiconductors with large Seebeck 
coefficients as potential thermoelectric materials.\cite{khazaei2012novel} However, up to
now these studies are mainly focused on the \ce{M_{2}X} 
and \ce{M_{3}X_{2}} type of MXenes, there are no
reports on the properties of exfoliated carbides or nitrides with 
other $n$.

In this work, we investigate systematically the evolution of  
structural and electronic properties of graphene-like titanium
carbide and nitride MXenes (\ce{Ti_{n+1}C_{n}} and
\ce{Ti_{n+1}N_{n}}, $n$ = 1--9) with different
functional groups (F, O, H, and OH) using conventional and hybrid
density functional calculations. We show
that all the studied materials are metals except for 
\ce{Ti_{2}CO_{2}}.
With the same surface termination, carbides and nitrides 
exhibit different electronic properties. The density of 
states (DOS) at the Fermi level increases in both carbides
and nitrides with distinct trends, as $n$ increases.
Moreover, the calculations suggest that the ``pure" carbide
and nitride monolayers are magnetic. 
After functionalization, magnetism is removed.

\section{\label{sec:comd}Computational details}
The first principle calculations are carried out using 
density functional theory (DFT)  as
implemented in the Vienna ab initio simulation
package (VASP)\cite{kresse1996efficient}. The 
all-electron projected argument wave (PAW) method\cite{blochl1994projector}
is used to describe the ion--electron interaction, with $1s^{1}$, $2s^{2}2p^{2}$,
$2s^{2}2p^{3}$, $2s^{2}2p^{4}$, $2s^{2}2p^{5}$, $3s^{2}3p^{1}$, and
$3d^{2}4s^{2}$ treated as valence electrons for H, C, N, O, F, Al, and Ti,
respectively. For Ti, PAW configurations including $3s$ and/or $3p$ 
electrons in the valence give the similar results. 
For the exchange-correlation
energy, we use both the Perdue--Burke--Ernzerhof
(PBE) version of generalized gradient approximation
(GGA)\cite{perdew1996generalized} and the Heyd--Scuseria--Ernzerhof 
(HSE06)\cite{heyd8207,paier2006screened,heyd2006erratum} 
hybrid functional. The latter can be described through the following
equation:
\begin{equation}
E_{xc}^{HSE}={1\over 4}E_{x}^{HF,SR}(\mu) + 
{3\over 4}E_{x}^{PBE,SR}(\mu) + 
E_{x}^{PBE,LR}(\mu) + E_{c}^{PBE}.
\end{equation}
The PBE exchange term is split into a short range (sr) 
and a long-range (lr) part. Then, 25\% of short range
part is replaced by a short range Hartree--Fock term.
The correlation part of PBE is not changed.
The screening
parameter in HSE was fixed at a value of 0.2 
\AA \ce{^{-1}}. This functional obtains improved formation energies
and more reliable band gaps than PBE,\cite{heyd8207,eyert2011vo} and is therefore better able to
identify stable structures and the metallic or insulating character of
the MXenes.

A plane wave cutoff energy
of 580 eV is sufficient to ensure the convergency of total energies to
1 meV per primitive cell.
The underlying structural
optimization are performed using 
the conjugated gradient method, and the 
convergency criterion was set to 10\ce{^{-6}}
eV/cell in energy and 0.005 eV/\AA  in force. A loose
12 $\times$ 12 $\times$ 1 Monkhorst--Pack $k$ grid 
is used during the optimization. To avoid any interaction 
between an MXene sheet and its periodically repeated images along
the $c$ axis, a large vacuum space of 20 \AA 
is used. After full structural optimization, a denser $k$ grid of 
42 $\times$ 42 $\times$ 1 and 20 $\times$ 20 $\times$ 1
is employed in the calculation of electronic properties for
PBE and HSE06, respectively. The computational cost of HSE06 precludes
use of the largest k-point grids.

\begin{figure}
\centering
\includegraphics[scale=1.5]{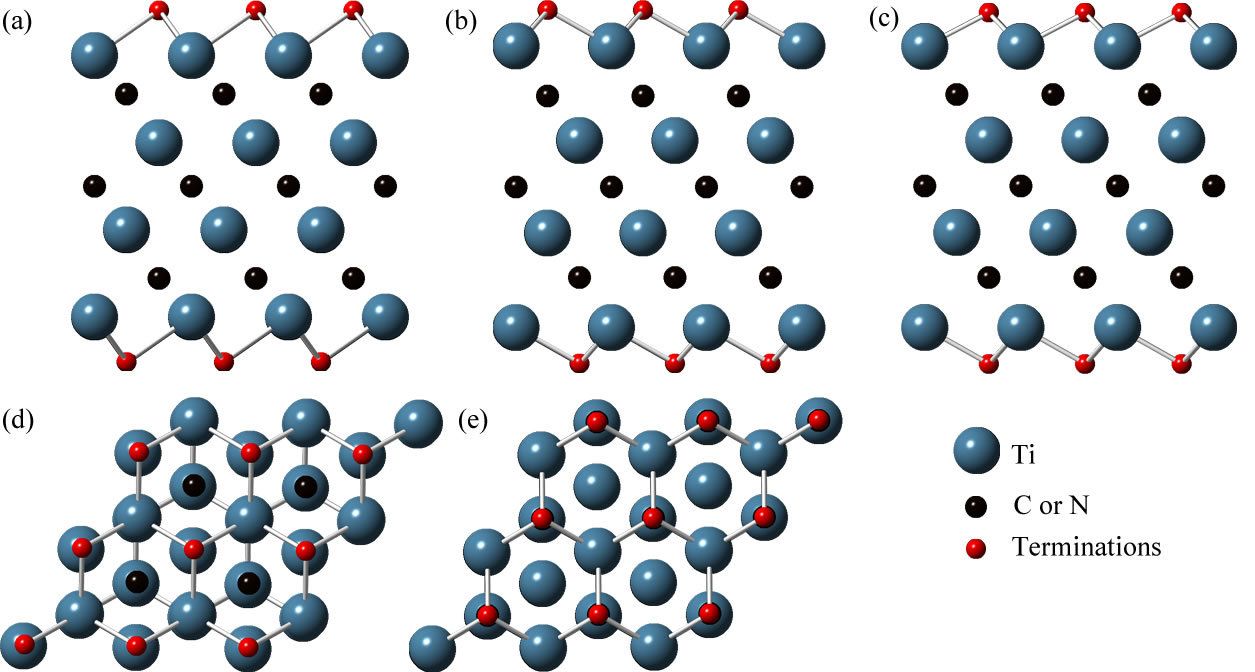}
\caption{(color online) Structure configurations of
functionalized MXenes with different arrangements of the surface atoms: side views of (a) I--\ce{Ti_{4}X_{3}T2}
, (b) II--\ce{Ti_{4}X_{3}T2}, and (c) III--\ce{Ti_{4}X_{3}T2}; (d)
and (e) top views of I--\ce{Ti_{4}X_{3}T2} and
II--\ce{Ti_{4}X_{3}T2}. Since configuration III is a mixture of 
I and II, the top view of III is not shown. }
\label{fig:config}
\end{figure}

\section{\label{sec:str}Structural models}
In order to examine the evolution of the electronic properties
of MXenes, we have studied the structural, magnetic, and
electronic properties of a series of carbides
\ce{Ti_{n+1}C_{n}} and nitrides 
\ce{Ti_{n+1}N_{n}}, with $n$ up to 9. 
The 2D MXenes are constructed appropriately 
by removing ``A" atoms
from their bulk MAX phases (Fig. \ref{fig:MAX}(b)). 
It is noteworthy that, for the parent MAX 
phases we used, the \ce{Ti_{2}AlC},\cite{manoun2006x}
\ce{Ti_{3}AlC_{2}},\cite{wang2010layered} \ce{Ti_{4}GeC_{3}},\cite{hogberg2005epitaxial}
\ce{(Ti, Nb)_{5}AlC_{4}},\cite{zheng2010ti0}
\ce{Ti_{7}SnC_{6}},\cite{zhang2009low} \ce{Ti_{2}AlN_{}},\cite{manoun2006x}
and \ce{Ti_{4}AlN_{3}}\cite{barsoum2000thermal} have been synthesized experimentally.
The unknown MAX phases (\ce{Ti_{n+1}AlC_{n}}, $n$ = 5, 7--9,
and \ce{Ti_{n+1}AlN_{n}}, $n$ = 2, 4--9) are built up by following a
generalized modular building principle proposed by
Etzkorn and co-workers.\cite{eklund2010m} 
The calculated lattice parameters of all studied 
MAX phases are shown in
Table \ref{MAX} and compared with available experimental
data\cite{manoun2006x,wang2010layered,hogberg2005epitaxial,
barsoum2000thermal,zheng2010ti0,zhang2009low} 
and previous theoretical results\cite{shein2012graphene,li2010first,keast2009prediction}.
The consistency of the results in both experiment and theory make us confident for further
studies.
\begingroup
\squeezetable
\begin{table}[b]
\caption{\label{MAX} Calculated PBE equilibrium lattice parameter ($a$, $c$, and
$c$/$a$) for MAX phases 
in comparison with available experimental data and previous theoretical results.
}
\begin{ruledtabular}
\begin{tabular}{lcccclcccc}
\ce{Ti_{n+1}AC_{n}} & & $a$ (\AA) &
$c$ (\AA) & $c$/$a$  &
\ce{Ti_{n+1}AN_{n}} & & $a$ (\AA)&
$c$ (\AA) & $c$/$a$ \\
\hline
\ce{Ti_{2}AlC_{}} & PBE & 3.0687 &	13.7266	& 4.47 &
\ce{Ti_{2}AlN_{}} & PBE & 2.9965&	13.6400	&4.55 \\
& Theo.\footnotemark[1] & 3.0694&	13.7311 &	4.47 & & Theo.\footnotemark[1] & 2.9965&	13.6360&	4.55 \\
& Exp.\footnotemark[2] & 3.065&	13.71&	4.48 & & Exp.\footnotemark[2] & 2.986	 & 13.60	&4.55 \\
\ce{Ti_{3}AlC_{2}} & PBE & 3.0816&	18.6379	&6.05 &
\ce{Ti_{3}AlN_{2}} & PBE & 3.0036	&18.5007&	6.16 \\
& Theo.\footnotemark[1] & 3.0824	&18.6522&	6.05 & & Theo.\footnotemark[1] & 3.0065	&18.4780	&6.15 \\
& Exp.\footnotemark[3] & 3.075&	18.58&	6.04 & &  & 	&&	 \\
\ce{Ti_{4}GeC_{3}} & PBE & 3.0868	&22.8515&	7.40 &
\ce{Ti_{4}AlN_{3}} & PBE & 2.9975	&23.4803	&7.83 \\
& Theo.\footnotemark[4] & 3.088	&22.852	&7.40 & & Theo.\footnotemark[1] & 2.9975	&23.4828	&7.83 \\
& Exp.\footnotemark[5] & 3.088	&22.852	&7.40 & & Exp.\footnotemark[6] & 2.9905	&23.380	&7.82 \\
\ce{Ti_{5}AlC_{4}} & PBE & 3.0791	&28.7440	&9.34 &
\ce{Ti_{5}AlN_{4}}  & PBE & 2.9928	&28.6249	&9.56 \\
& Theo.\footnotemark[7] & 3.074	&28.76	&9.35 & & Theo.\footnotemark[8] & 3.00	&28.44	&9.48 \\
\ce{(Ti,Nb)_{5}AlC_{4}} & Exp.\footnotemark[7] & 3.100	&28.89	&9.32 & & & & & \\
\ce{Ti_{6}AlC_{5}} & PBE & 3.0841	&33.5105&	10.86 &
\ce{Ti_{6}AlN_{5}} &PBE  & 3.0011&33.3420		&11.11 \\
\ce{Ti_{7}SnC_{6}} & PBE & 3.1604	&40.5154	&12.82 &
\ce{Ti_{7}AlN_{6}} & PBE & 3.0017	&38.2657	&12.75 \\
& Exp.\footnotemark[9] & 3.200&	41.000	&12.81 & &  & 	&	& \\
\ce{Ti_{8}AlC_{7}} & PBE & 3.0817	&43.4258	& 14.09 &  
\ce{Ti_{8}AlN_{7}}  & PBE &3.0036 &43.1310		& 14.36\\
\ce{Ti_{9}AlC_{8}} & PBE & 3.0878	&48.1608	& 15.59 &  
\ce{Ti_{9}AlN_{8}}  &  PBE & 3.0132&	47.7686	& 15.85 \\
\ce{Ti_{10}AlC_{9}} & PBE & 3.0885	&53.1124	&17.19&  
\ce{Ti_{10}AlN_{9}} &  PBE &3.0145 &		52.6603& 17.47 \\
\end{tabular}
\end{ruledtabular}
\footnotetext[1]{Ref.~\onlinecite{shein2012graphene}, VASP--PBE.}
\footnotetext[2] {Ref.~\onlinecite{manoun2006x}, XRD.}
\footnotetext[3] {Ref.~\onlinecite{wang2010layered}, XRD.}
\footnotetext[4] {Ref.~\onlinecite{li2010first}, DFT--GGA.}
\footnotetext[5] {Ref.~\onlinecite{hogberg2005epitaxial}, XRD.}
\footnotetext[6] {Ref.~\onlinecite{barsoum2000thermal}, XRD.}
\footnotetext[7] {Ref.~\onlinecite{zheng2010ti0}, experiment, XRD/STEM; theory, CASTEP--GGA.}
\footnotetext[8] {Ref.~\onlinecite{keast2009prediction}, FPLAPW--GGA.}
\footnotetext[9] {Ref.~\onlinecite{zhang2009low}, XRD/DSC/HRTEM.}
\end{table}
\endgroup
Notably, it has been reported that
the surfaces of MXenes are covered with oxygen-containing
groups, such as OH, and fluorine, F.\cite{naguib2011two} 
As a consequence, 
the chemistry of exfoliated MXenes is expected to be closer to
the functionalized MXenes than the bare material.
It also could be possible to create
O-termination surfaces by post-processing of OH-terminated
systems.\cite{khazaei2012novel}
Moreover, the hydrogen-terminated graphene has been
studied extensively due to its unique properties.\cite{hou2011effect} 
Therefore,
in our studies, we have considered that the surfaces of 
MXenes are fully terminated by F, O, H, and OH, with
a general formula of \ce{Ti_{n+1}X_{n}T_{2}} ( 
X = C, N, and T is the terminations).
Note, for MXenes extracted from the unknown MAX phases,
only the fluorine termination has been studied.
According to the geometry of the MXenes, the functionalized
MXenes are built up with three major possible configurations: (I) 
all the functional groups are located above the hollow site of
three neighboring C/N atoms and pointed 
to the Ti atoms in the second Ti atomic layer on both sides
of MXenes; (II) all the functional groups are located on
top the topmost sides of C/N atoms on both sides of MXenes;
(III) is a combination of (I) and (II), in which one functional
group is located on top of the hollow sites of C/N atoms on
one side and another one locates above the top sites of C/N
atoms on other sides. These models are shown in Fig. \ref{fig:config}.
We perform full geometry optimization
for each of these structures.

\section{Results and discussion}
\subsection{Structural properties}

\begingroup
\squeezetable
\begin{table}[b]
\caption{\label{conf}PBE total energy differences (relative to configuration I) 
of functionalized \ce{Ti_{n+1}C_{n}} and 
\ce{Ti_{n+1}C_{n}} MXenes. 
The most stable configuration is highlighted in bold-typeface.}
\begin{ruledtabular}
\begin{tabular}{lcccclcccc}
\multirow{2}{*}{\ce{Ti_{n+1}C_{n}}} & F & O & H & OH & 
\multirow{2}{*}{\ce{Ti_{n+1}N_{n}}}& F & O & H & OH \\
\cline{2-5}\cline{7-10}
 & \multicolumn{4}{c}{\ce{E_{0}-E_{0}^{I}} (eV)}
&  & \multicolumn{4}{c}{\ce{E_{0}-E_{0}^{I}} (eV)} \\
\hline
\ce{Ti_{2}C} & & & & & \ce{Ti_{2}N} \\
I & \textbf{0.000} & \textbf{0.000} & \textbf{0.000} & \textbf{0.000} &
I & \textbf{0.000} & \textbf{0.000} & \textbf{0.000} & \textbf{0.000} \\
II & 0.439& 1.750&0.698 &0.321 &
II & 0.398 & 1.175 & 0.533 & 0.296 \\
III & 0.215& 0.769 &0.365 &0.130 &
III & 0.427 & 0.546 & 0.299 & 0.266 \\
\ce{Ti_{3}C_{2}} & & & & & \ce{Ti_{3}N_{2}} \\
I & \textbf{0.000} & \textbf{0.000} & \textbf{0.000} & \textbf{0.000} &
I & 0.000 & \textbf{} & \textbf{} & \textbf{} \\
II & 0.729& 1.423&0.718 &0.476 &
II & \textbf{-0.089} &  &  &  \\
III & 0.371& 0.712&0.349 &0.240 &
III & -0.002 &  &  &  \\
\ce{Ti_{4}C_{3}} & & & & & \ce{Ti_{4}N_{3}} \\
I & \textbf{0.000} & \textbf{0.000} & \textbf{0.000} & \textbf{0.000} &
I & \textbf{0.000} & \textbf{0.000} & \textbf{0.000} & \ 0.000 \\
II & 0.720& 1.462&0.719 &0.455 &
II & 0.161 & 1.102 & 0.591 & 0.003 \\
III & 0.387& 0.733&0.361 &0.245 &
III & 0.088 & 0.561 & 0.299 & \textbf{-0.003} \\
\ce{Ti_{5}C_{4}} & & & & & \ce{Ti_{5}N_{4}} \\
I & \textbf{0.000} & \textbf{0.000} & \textbf{0.000} & \textbf{0.000} &
I & \textbf{0.000} & & & \\
II & 0.701& 1.491&0.720 &0.429 &
II & 0.051 & & & \\
III & 0.362& 0.747&0.358 &0.222 &
III & 0.057 & & & \\
\ce{Ti_{6}C_{5}} & & & & & \ce{Ti_{6}N_{5}} \\
I & \textbf{0.000} & & & &
I & \textbf{0.000} & & & \\
II & 0.720& & & &
II & 0.146 & & & \\
III & 0.369 & & & &
III & 0.08 & & & \\
\ce{Ti_{7}C_{6}} & & & & & \ce{Ti_{7}N_{6}}  \\
I & \textbf{0.000} & \textbf{0.000} & 0.000 & \textbf{0.000} &
I & \textbf{0.000} & & & \\
II & 0.712& 1.532&0.205 &0.419 &
II & 0.093 & & & \\
III & 0.361& 0.780& \textbf{-0.154} &0.213 &
III & 0.082 & & & \\
\ce{Ti_{8}C_{7}} & & & & & \ce{Ti_{8}N_{7}} \\
I & \textbf{0.000} & & & &
I & \textbf{0.000} & & & \\
II & 0.728& & & &
II & 0.125 & & & \\
III & 0.371& & & &
III & 0.072 & & & \\
\ce{Ti_{9}C_{8}} & & & & & \ce{Ti_{9}N_{8}} \\
I & \textbf{0.000} & & & &
I & \textbf{0.000} & & & \\
II & 0.728& & & &
II & 0.125 & & & \\
III & 0.368& & & &
III & 0.079 & & & \\
\ce{Ti_{10}C_{9}} & & & & & \ce{Ti_{10}N_{9}} \\
I & \textbf{0.000} & & & &
I & \textbf{0.000} & & & \\
II & 0.729& & & &
II & 0.127 & & & \\
III & 0.367& & & &
III & 0.074 & & & \\
\end{tabular}
\end{ruledtabular}
\end{table}
\endgroup

We first check the relative stability of the different structural models.
The PBE total energy differences (relative to configuration I) 
of all configurations and
terminations are listed in Table \ref{conf}. 
Configuration I usually has the lowest total energy indicating it is 
energetically more favorable and the configuration III normally has
the highest energy. This is quite understandable, since Ti atoms are the main
electron donors and the functional groups are more likely to bond with
Ti atoms than C or N atoms. However, there are several exceptions. 
For \ce{Ti_{3}N_{2}F_{2}}, configuration II has the lowest energy;
whereas for \ce{Ti_{4}N_{3}(OH)_{2}}
and \ce{Ti_{7}C_{6}H_{2}},
the configuration III is more stable. We also notice that the relative
total energy differences are not very sensitive to increased layer
thickness and oxygen always has the largest differences among the four
functional groups for both carbides and nitrides. Moreover, the energy differences
of nitrides are smaller than carbides. 
In general, our results are consistent
with previous reports.\cite{khazaei2012novel,tang2012mxenes}

\begingroup
\squeezetable
\begin{table}[b]
\caption{\label{stru} Calculated equilibrium lattice parameter ($a$),
thickness of the monolayer ($L$),
and bond lengths between surface Ti atom and termination (\ce{d_{Ti-T}}),
and nearest C/N atom (\ce{d_{Ti-X}}) of selected functionalized MXenes
in comparison with previous theoretical results.
}
\begin{ruledtabular}
\begin{tabular}{lccccclccccc}
\ce{Ti_{n+1}C_{n}T_{2}} & & $a$ (\AA) & $L$ (\AA) &
\ce{d_{Ti-C}} (\AA) & \ce{d_{Ti-T}} (\AA) &
\ce{Ti_{n+1}N_{n}T_{2}} & & $a$ (\AA)& $L$ (\AA) &
\ce{d_{Ti-N}} (\AA) & \ce{d_{Ti-T}} (\AA) \\
\hline
\ce{Ti_{2}C_{}F_{2}} & Theo.\footnotemark[1] & 3.0587 &4.84&	2.10	& 2.16 &
\ce{Ti_{2}N_{}F_{2}} & Theo.\footnotemark[1] & 3.0704&4.69&	2.07	&2.16 \\
& PBE & 3.0583&	4.80&2.10 &	2.16 & & PBE & 3.0642&4.63&	2.07&	2.16 \\
& HSE & 3.0312&4.72&	2.08&	2.14 & & HSE & 3.0460	 &4.55& 2.05	&2.14 \\
\ce{Ti_{2}C_{}O_{2}} & Theo.\footnotemark[1] & 3.0349&4.47&	2.19	&1.98 &
\ce{Ti_{2}N_{}O_{2}} & Theo.\footnotemark[1] & 3.0191	&4.28&2.17&	1.93 \\
& PBE & 3.0329&4.45	&2.19&	1.98 & & PBE & 3.0030	&4.35&2.15	&1.98 \\
& HSE & 3.0093&	4.40&2.17&	1.96 & & HSE & 2.9743	&4.29&2.13&	1.86 \\
\ce{Ti_{2}C_{}(OH)_{2}} & Theo.\footnotemark[1] & 3.0715	&6.75&2.11&	2.18 &
\ce{Ti_{2}N_{}(OH)_{2}} & Theo.\footnotemark[1] & 3.0635	&6.68&2.08	&2.17 \\
& PBE & 3.0712	&6.79&2.12	&2.18 & & PBE & 3.0635	&6.66&2.08	&2.17 \\
& HSE & 3.0464	&6.71&2.09	&2.17 & & HSE & 3.0447	&6.56&2.06	&2.16 \\
\ce{Ti_{2}C_{}H_{2}} & PBE & 3.0357	&4.28&2.10	&2.01 &
\ce{Ti_{2}N_{}H_{2}}  & PBE & 2.9882	&4.20&2.07	&2.00 \\
& HSE & 3.0094	&4.24&2.08	&1.99 & & HSE & 3.0127	&4.11&2.04	&2.00 \\
\ce{Ti_{3}C_{2}F_{2}} & Theo.\footnotemark[1] & 3.0792	&7.23&2.08&	2.17 &
\ce{Ti_{3}N_{2}F_{2}} &  & &		& \\
& PBE & 3.0775	&7.21&2.08	&2.17 & & PBE & 3.0226&	7.15&2.07&	2.15 \\
& HSE & 3.0508&	7.14&2.06	&2.15 & & HSE & 2.9945	 & 7.07& 2.04	& 2.14 \\
\ce{Ti_{4}C_{3}F_{2}} & Theo.\footnotemark[1] & 3.0856	&9.70&2.08	&2.17 &
\ce{Ti_{4}N_{3}F_{2}} & Theo.\footnotemark[1] & 3.0287	&9.45&2.06	&2.15 \\
& PBE & 3.0844&	9.67&2.07	&2.17 & & PBE & 3.0268	&9.62&2.06	&2.15 \\
& HSE & 3.0582&9.58&	2.05	&2.16 & & HSE & 3.0041	&9.49&2.04	&2.14 \\
\ce{Ti_{5}C_{4}F_{2}} & Theo.\footnotemark[1] & 3.0828	&12.15&2.08	& 2.17 &  
\ce{Ti_{5}N_{4}F_{2}}  &  & &		& \\
& PBE & 3.0794	&12.19&2.08	&2.17& & PBE & 3.0226&	12.09&2.07&	2.15 \\
& HSE & 3.0528	&12.09&2.06	&2.15 & & HSE & 2.9947	 & 11.96&2.04	&2.13 \\
\ce{Ti_{6}C_{5}F_{2}} & PBE & 3.0744	&14.72&2.08	& 2.17 &  
\ce{Ti_{6}N_{5}F_{2}}  &  PBE & 3.0235&14.52&	2.05	& 2.16 \\
& HSE & 3.0456	&14.63&	2.06&2.15 & & HSE & 	2.9932 & 14.37&2.04	&2.13 \\
\ce{Ti_{7}C_{6}F_{2}} & PBE & 3.0719	&17.24&2.08	&2.17&  
\ce{Ti_{7}N_{6}F_{2}} &  PBE &3.0222 &16.97&		2.05& 2.15 \\
& HSE & 3.0441	&17.13&2.06	&2.15 & & HSE & 2.9945	 &16.78& 2.03	& 2.14\\
\ce{Ti_{8}C_{7}F_{2}} & PBE & 3.0730	&19.73&2.08	& 2.17 &  
\ce{Ti_{8}N_{7}F_{2}}&  PBE & 3.0194&19.45&	2.05	& 2.15 \\
& HSE & 3.0442	&19.61&2.08	& 2.15& & HSE & 2.9913	 & 19.23&2.03	& 2.13\\
\ce{Ti_{9}C_{8}F_{2}} & PBE & 3.0728	&22.22&2.08	& 2.17 &  
\ce{Ti_{9}N_{8}F_{2}}  &  PBE & 3.0200&21.87&	2.05	& 2.15 \\
& HSE & 3.0453	&	22.08&2.06&2.15 & & HSE & 2.9932	 & 21.62&2.03	&2.13 \\
\ce{Ti_{10}C_{9}F_{2}}& PBE & 3.0723	&24.72&2.08	& 2.17 &  
\ce{Ti_{10}N_{9}F_{2}}  &  PBE & 3.0179&24.34&	2.05	& 2.15 \\
& HSE & 3.0448	&24.56&2.06	&2.15 & & HSE & 2.9874	 & 24.13&2.03	&2.13 \\
\end{tabular}
\end{ruledtabular}
\footnotetext[1]{Ref.~\onlinecite{khazaei2012novel},
VASP--PBE.}
\end{table}
\endgroup

We then proceed to investigate the ground-state structural
properties of the functionalized MXenes. Some of  the calculated
equilibrium lattice parameters $a$, thickness of the monolayer
(vertical distance from the topmost atomic layer to bottommost atomic layer, $L$),
 and bond lengths between
surface Ti atom and termination (\ce{d_{Ti-T}}) and nearest 
C/N atom (\ce{d_{Ti-X}}) with PBE and HSE06 functionals
are shown and compared with other theoretical results\cite{khazaei2012novel} 
in Table \ref{stru}. Clearly, our PBE results are in good agreement 
with others. The HSE06 lattice parameters and bond lengths are about
0.02 $\sim$ 0.03 \AA\  \ce{smaller} than for PBE, except for 
\ce{Ti_{2}N_{}H_{2}}, where the HSE06 lattice
parameter is 0.15 \AA\  longer. For both carbides and nitrides,
the oxygen terminated MXenes have the shortest \ce{d_{Ti-T}}
and longest \ce{d_{Ti-X}} bond length, implying a strong interaction
between surface Ti atoms and O termination. Alternatively,
the hydroxyl terminated MXenes have the largest lattice parameters
and longest \ce{d_{Ti-T}} bond lengths. Comparing carbides with
nitrides, we find that the carbides have larger lattice constant and longer bond
lengths, correlating with the atomic radii difference between
carbon and nitrogen. With increasing $n$, taking fluorinated MXenes
as an example, the lattice constant of carbides increases initially, then
decreases slowly when $n$ $>$ 5. On the other hand, the lattice
constant of nitrides drops rapidly at $n$ = 2, then it remains 
the same around 3.0 \AA. For MXenes covered with other
functional groups, similar behaviors have been observed.

\subsection{Magnetic and electronic properties}

\begin{figure}[b]
\centering
\includegraphics[scale=0.7]{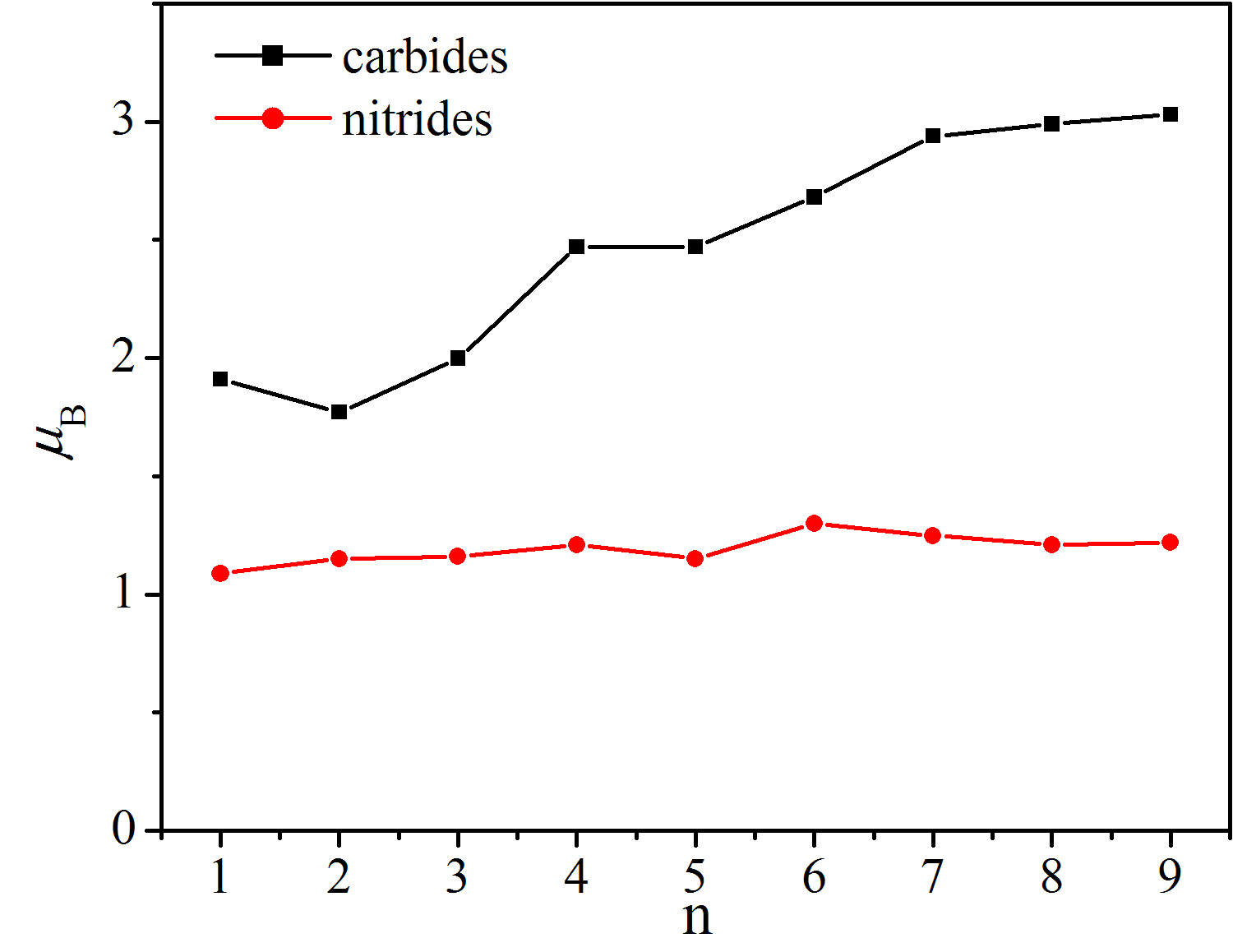}
\caption{(color online) Evolution of total magnetic moment of
pure carbide (black line) and nitride (red line) monolayers as a function
of the layer thickness. Carbides and nitrides behave differently.}
\label{fig:mage}
\end{figure}

We checked for possible magnetic ground states of MXenes with
fully relaxed spin--polarized calculations. We find that only the
\ce{Ti_{n+1}X_{n}} monolayers are magnetic. The
magnetization results mainly from the $3d$ electrons of surface
Ti atoms. Taking \ce{Ti_{2}X_{}} as an example, the magnetic moments
of  surface Ti and topsite C/N atoms are about 0.982 and 0.065 
\ce{\mu_{B}} per atom for \ce{Ti_{2}C} and 0.619 and 0.027
\ce{\mu_{B}} per atom for \ce{Ti_{2}N}, respectively.
The total magnetic moment of carbides and nitrides as a function of
$n$ are presented 
in Fig. \ref{fig:mage}. As we can see, carbides and nitrides
show a different behavior. The total magnetic moment of carbides
increases from 2 to 3 \ce{\mu_{B}}, while in nitrides the
total magnetic moment fluctuates around 1.2 \ce{\mu_{B}}.
Upon functionalization, the magnetism of MXenes disappeared.

\begin{figure}[b]
\centering
\includegraphics[scale=1.5]{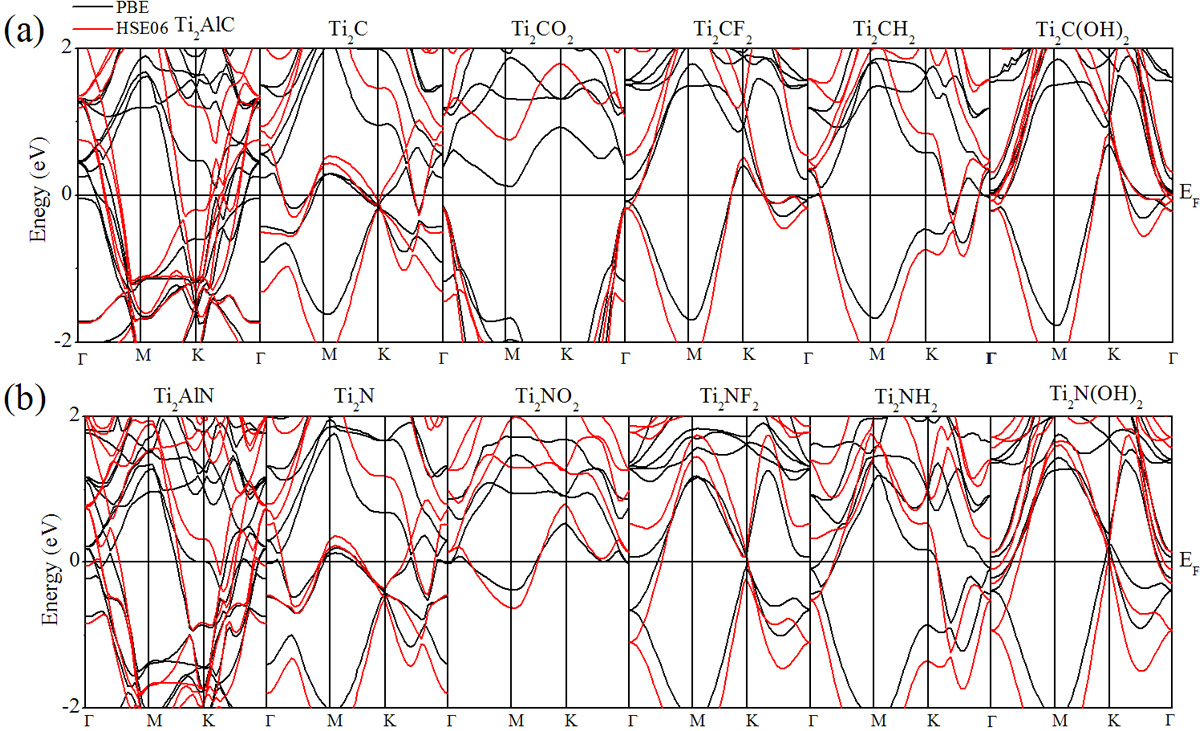}
\caption{(color online) (a) and (b) Band structures of \ce{Ti_{2}C_{}T_{2}}
and \ce{Ti_{2}N_{}T_{2}}, and related MXenes and MAX phases. 
The HSE06 results are
similar to PBE results, with modest changes to the band structures. The band-gap of
\ce{Ti_{2}C_{}O_{2}} is widened by HSE06.}
\label{fig:bs}
\end{figure}

With the appropriate ground-states, we can now move to
study the electronic properties of functionalized MXenes.
We first examine the electronic properties of the thinnest
\ce{Ti_{2}X_{}T_{2}} monolayers.
The calculated PBE and 
HSE06 band structures of  \ce{Ti_{2}X_{}T_{2}} 
monolayers are shown and compared with pure MXenes and parent 
MAX phases  in Fig. ~\ref{fig:bs}. The PBE and HSE06 functionals
give the similar results, and all of these materials
are metals, with the notable exception of \ce{Ti_{2}C_{}O_{2}}. 
PBE functional predicts a band gap of 0.24
eV for \ce{Ti_{2}C_{}O_{2}}, while the gap is widened to
0.88 eV by HSE06 hybrid functional. Since HSE06 is expected to be more
reliable than PBE for band gaps, we predict  a near 1eV band gap for
this material. Unless specifically mentioned, the following
results are all calculated with HSE06.

\begin{figure}
\centering
\includegraphics[scale=1.1]{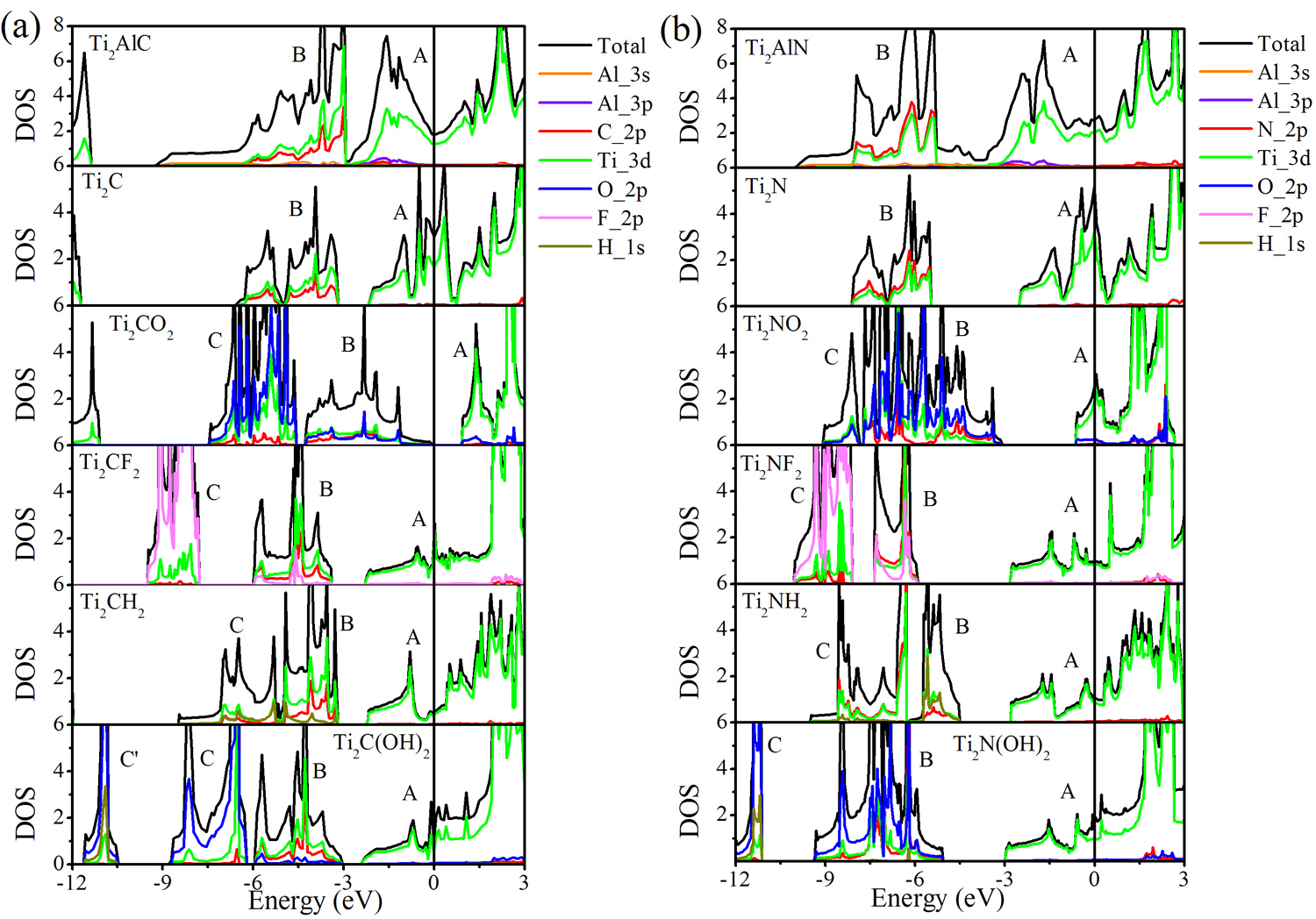}
\caption{(color online) (a) and (b) Partial density of states of \ce{Ti_{2}C_{}T_{2}}
and \ce{Ti_{2}N_{}T_{2}}, and related MXenes and MAX phase computed
using the HSE06 functional. }
\label{pdos}
\end{figure}

By examining the band structures near
the Fermi level, we find that the Fermi energy shifts down after removing
Al atoms from MAX phases, and then it shifts down in energy again after the MXenes surfaces
are terminated (Fig. \ref{fig:bs}). To give a better description of the electronic properties, the partial 
density of states (PDOS) of these materials are shown in Fig. \ref{pdos}.
As we can see, for MAX phases, the DOS at Fermi level (\ce{N(E_{F})}) is dominated by
Ti $3d$ orbitals. The valence states below Fermi level can be divided into
two sub-bands which are formed by hybridized Ti $3d$--C $2p$ and 
Ti $3d$--Al $3s$ orbitals between -10 and -3 eV (band B), and Ti $3d$--Al $3p$ 
orbitals near Fermi level (band A). These mixed states correspond to 
the Ti--C and Ti--Al bands. By extracting Al atoms, the bands A and B are 
narrowed, and a gap opens between them. The \ce{N(E_{F})} increases 
from 1.88 to 3.15 and from 2.77 to 4.84 for carbides and nitrides, due to the
breaking of Ti--Al bonds, respectively. After functionalization, the Fermi level
shifts to lower energy states and the \ce{N(E_{F})} decreases due to the
new energy states between Ti and terminations. For carbides, band C is formed below band B
corresponding to the hybridization between Ti and functional groups. Different
from carbides, except for the newly formed band C, 
the functional groups also have a large contribution for band B in nitrides.
Comparing \ce{Ti_{2}C_{}O_{2}} with other materials, we find that
band B is equally contributed by Ti $3d$, C $2p$, and O $2p$ orbitals.
The strong hybridization of Ti $3d$--C $2p$ and Ti $3d$--O $2p$
is responsible for the semiconducting behavior of \ce{Ti_{2}C_{}O_{2}}.

\begingroup
\squeezetable
\begin{table}
\caption{\label{DOS}Evolution of DOS at Fermi level
as a function of $n$ of functionalized MXenes computed
using the HSE06 functional.
}
\begin{ruledtabular}
\begin{tabular}{lcccclcccc}

\ce{Ti_{n+1}C_{n}} & O & F & H & OH
& \ce{Ti_{n+1}C_{n}} & O & F & H & OH \\
\hline
\ce{Ti_{2}C_{}} & 0& 2.156& 0.566& 1.916& \ce{Ti_{2}N_{}} & 2.562 & 0.941&1.045&1.817 \\
\ce{Ti_{3}C_{2}} & 0.575& 2.026& 0.823& 2.876& \ce{Ti_{3}N_{2}} &  & 1.279&& \\
\ce{Ti_{4}C_{3}} & 0.98& 2.013& 2.902& 3.176& \ce{Ti_{4}N_{3}} &3.895  & 2.601&2.313& 3.881\\
\ce{Ti_{5}C_{4}} & 0.915& 4.392& 5.124& 4.013& \ce{Ti_{5}N_{4}} &  & 4.895&& \\
\ce{Ti_{6}C_{5}} & & 7.284& & &\ce{Ti_{6}N_{5}} &  & 3.407&& \\
\ce{Ti_{7}C_{6}}& 0.998& 6.579& 8.082&6.214 &\ce{Ti_{7}N_{6}} &  & 6.501&& \\
\ce{Ti_{8}C_{7}} & & 5.835& & & \ce{Ti_{8}N_{7}} &  & 4.072&& \\
\ce{Ti_{9}C_{8}} & & 5.444& & & \ce{Ti_{9}N_{8}} &  & 7.363&& \\
\ce{Ti_{10}C_{9}}& & 5.248& & & \ce{Ti_{10}N_{9}} &  & 8.107&& \\
\end{tabular}
\end{ruledtabular}
\end{table}
\endgroup

\begin{figure}
\centering
\includegraphics[scale=1]{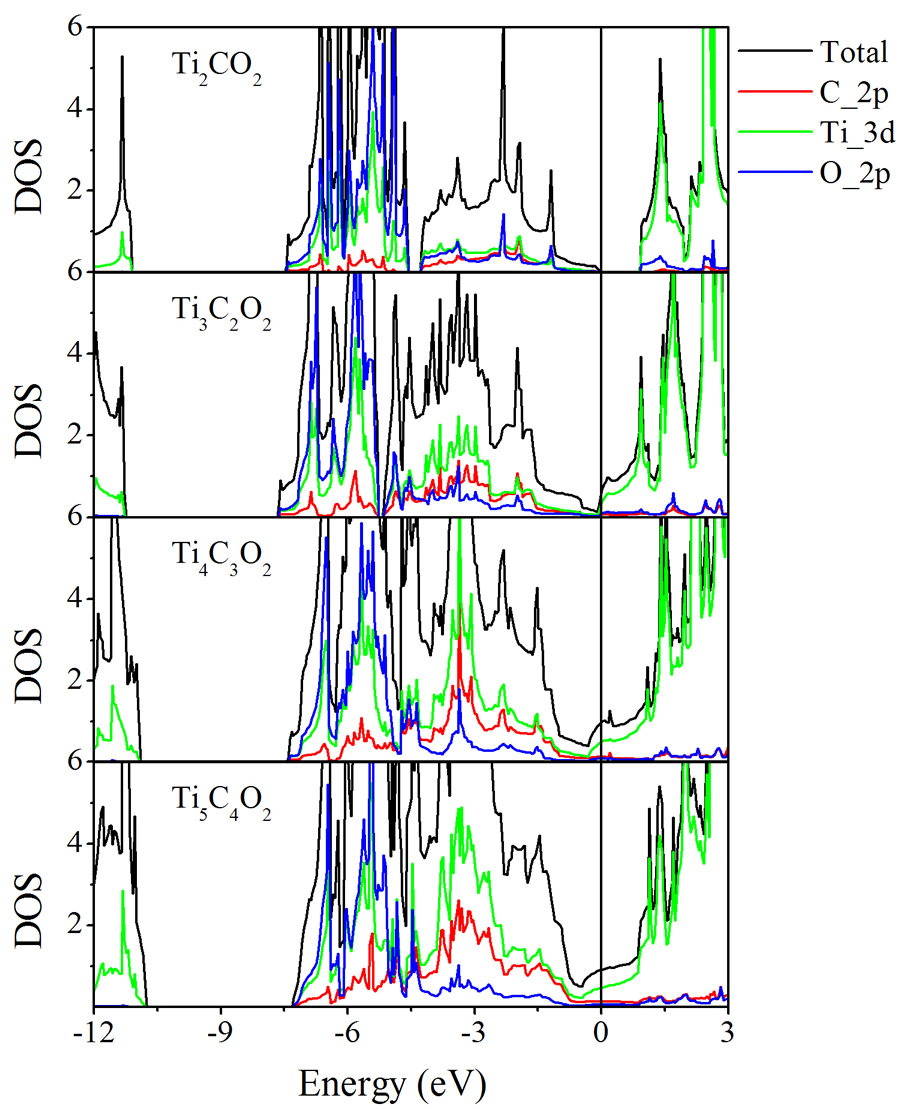}
\caption{(color online) Partial density of states of \ce{Ti_{n+1}C_{n}O_{2}}
up to $n$ = 4 computed using the HSE06 functional.}
\label{oxygen}
\end{figure}

Since the \ce{N(E_{F})} is very important for the surface chemistry of
layered materials, the evolution of the value of \ce{N(E_{F})} of 
functionalized MXenes  as a function of $n$ has been studied
and is shown in Table ~\ref{DOS}. 
Firstly, we notice that starting
from $n$ = 2, the carbides with oxygen termination are metals.
By looking at the PDOS \ce{Ti_{n+1}C_{n}O_{2}}
(Fig. ~\ref{oxygen}), it is clear that the contribution of O $2p$ orbitals
to band B decreases as $n$ increased. Starting from $n$ = 2, Band B is dominated by mixed
Ti $3d$ and C $2p$ bands and connected with band A , 
which is similar to the case observed in MAX phases. Therefore, the 
weaker Ti--O coupling in energy states B results in the metallic phase of 
\ce{Ti_{n+1}C_{n}O_{2}}.
Secondly, we find that, for carbides, the fluorine terminated MXenes have the largest DOS at 
Fermi level and the oxygen terminations have the lowest one, while it is 
opposite for nitrides. Thirdly, there is a clear trend that, for both carbides
and nitrides covered with functional groups, the value of \ce{N(E_{F})} increases 
with increased $n$. For instance, taking fluorinated MXenes as an example,
\ce{N(E_{F})} of \ce{Ti_{n+1}C_{n}F_{2}} is around 2 for $n$ $\leqslant$
3, then it increases rapidly from 2.013 ($n$ = 3) to
7.284 ($n$ = 5) , which is about 3.5 times higher. It decreases slowly 
to 5.248 ($n$ = 9), which is still 2.5 times higher than the thinnest MXenes.
Alternatively, \ce{N(E_{F})} of \ce{Ti_{n+1}N_{n}F_{2}} continuously
increases from 0.941 ($n$ = 1) to 8.107 ($n$ = 9), which is about 8 times higher.
The similar behavior is also observed in MXenes with other functional groups.
The huge difference of \ce{N(E_{F})} between the thin ($n$ $<$ 5) and thick
($n$ $\geqslant$ 5) functionalized MXenes implies their surface properties should be 
very different and the thick layers may have higher chemical
activity. The conductivity is expected to be similarly effected. 

\begin{figure}
\centering
\includegraphics[scale=1]{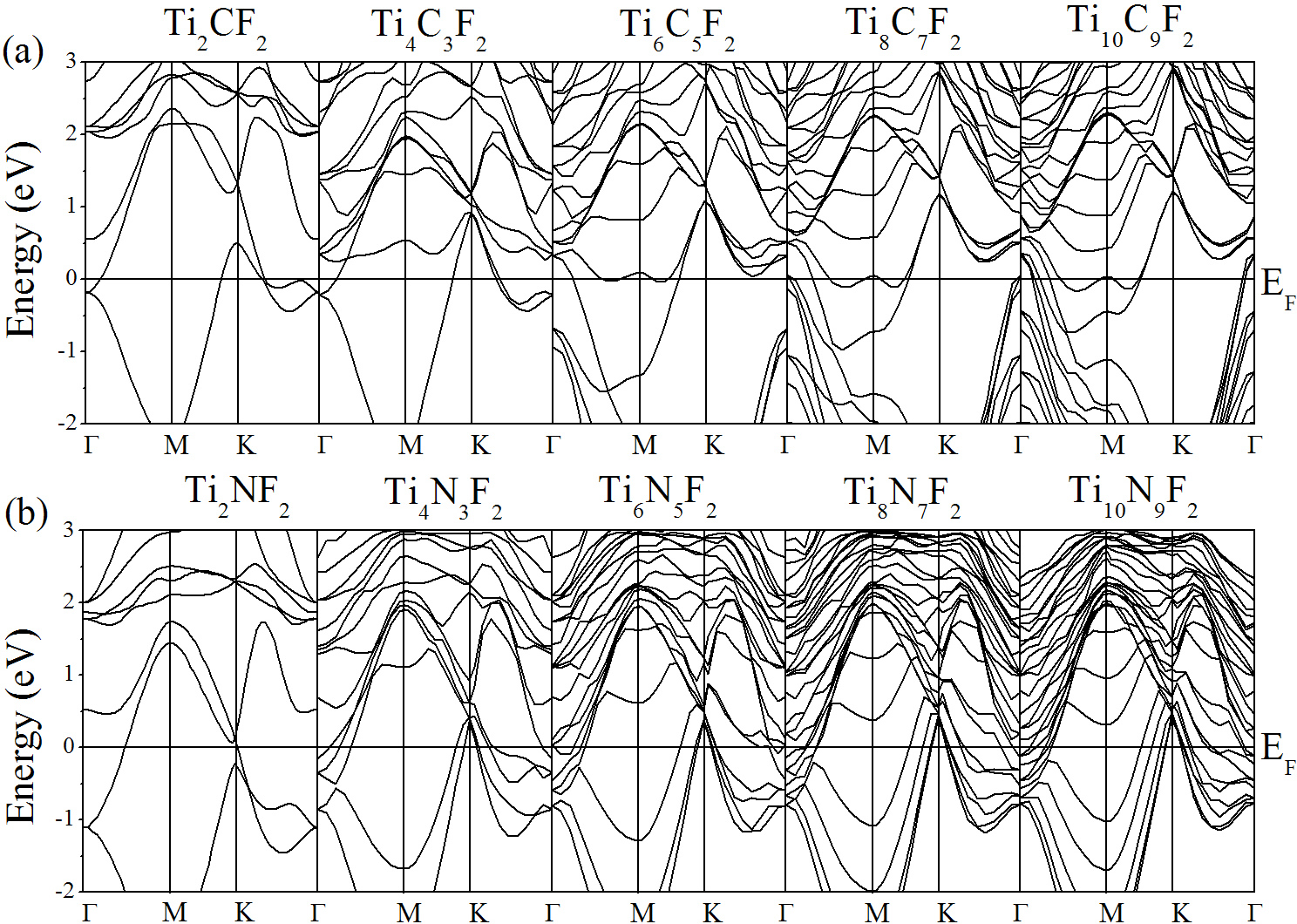}
\caption{(color online) (a) and (b) Band structures of 
\ce{Ti_{n+1}C_{n}F_{2}} and \ce{Ti_{n+1}N_{n}F_{2}} at
$n$ = 1, 3, 5, 7, and 9 computed using the HSE06 functional. Clearly, there are more bands forming
near the Fermi level with increased $n$.}
\label{tband}
\end{figure}

To better understand the trend, we examine the PDOS of  \ce{Ti_{n+1}X_{n}F_{2}} 
as a function of $n$ as
shown in Fig. ~\ref{pdos} and compare it with bulk titanium carbide (TiC) and nitride (TiN).
\begin{figure}
\centering
\includegraphics[scale=1]{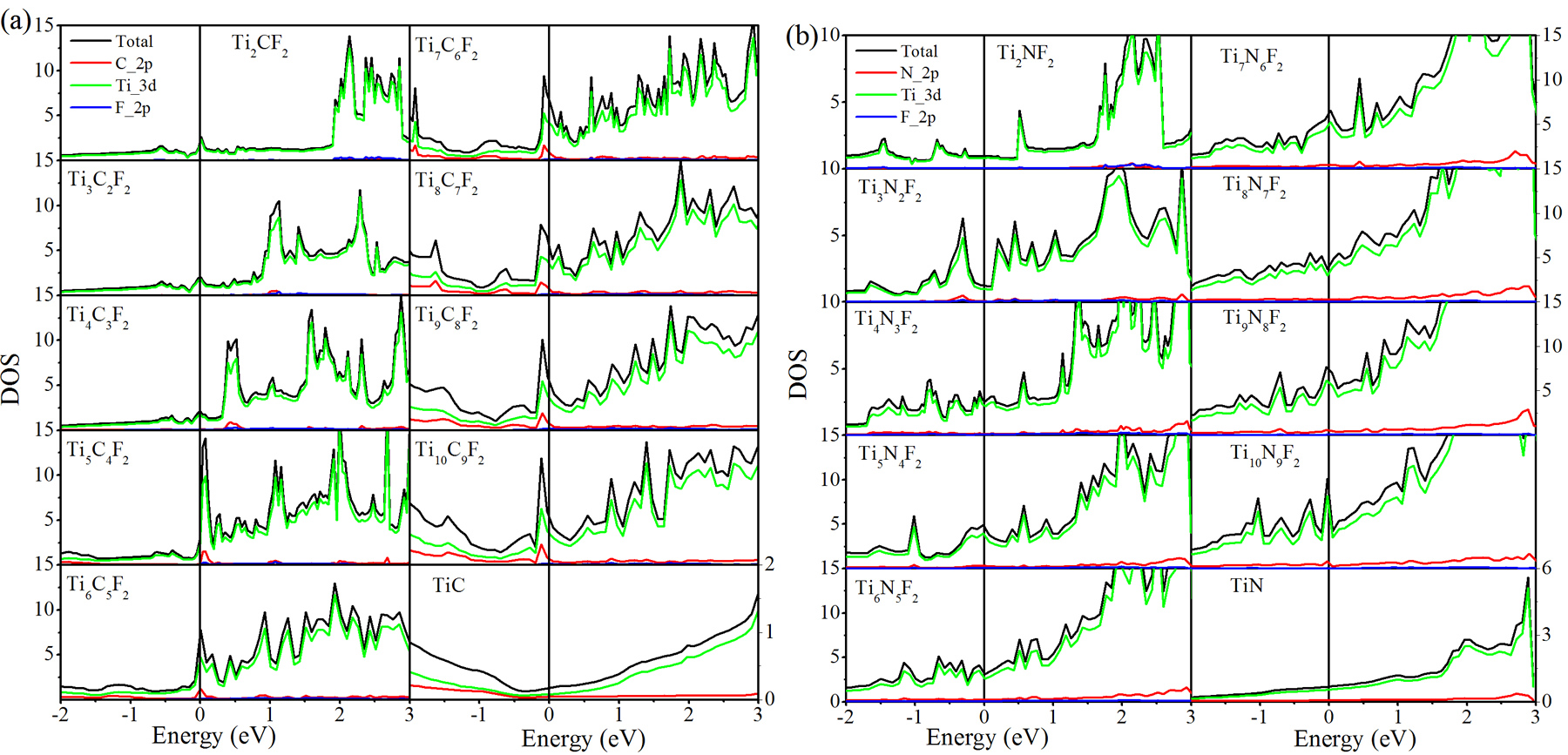}
\caption{(color online) (a) and (b) Partial density of states of 
\ce{Ti_{n+1}C_{n}F_{2}} and \ce{Ti_{n+1}N_{n}F_{2}}, compared
with bulk TiC and TiN phase, computed
using the HSE06 functional.}
\label{pdos2}
\end{figure}
One can observe that at $n$ = 1, the peak of the conduction states is about 2 eV above
the Fermi level for both carbide and nitride. This can also be seen from the band
structures (Fig. ~\ref{tband}). We also find that, for carbide, there are energy bands
crossing the Fermi level along all the high symmetry directions of the Brillouin Zone. 
However, the energy
bands only cross the $\Gamma$--M and K--$\Gamma$ directions in nitride.
This could be the reason that \ce{N(E_{F})} of \ce{Ti_{2}N_{}F_{2}}
is about half of \ce{Ti_{2}C_{}F_{2}}. With increasing $n$, the Fermi level
shifts down continuously and there are more bands forming near the Fermi
level (Fig. ~\ref{tband}) related to the Ti--C coupling (Fig. ~\ref{pdos2}). 
For carbides, we observe a sharp narrow DOS peak appears around 0.5 eV 
above the 
Fermi level at $n$ = 3, corresponding to the less dispersed band along the
$\Gamma$--M--K direction (Fig. ~\ref{tband}(a)). 
This peak shifts down and come across the 
Fermi level at $n$ = 4, where we start to see the increasing of \ce{N(E_{F})}. 
The latter
reaches the maximum when Fermi level locates at the top of this peak at
$n$ = 5. For higher $n$, this band keeps shifting down and goes below the
Fermi level (Fig. ~\ref{tband}(a)). Moreover, there are fewer bands
cross the Fermi level in the K--$\Gamma$ direction. Although a new band
is approaching the Fermi level, the value of \ce{N(E_{F})}
is still decreased.
For nitrides, we don't see the sharp narrow DOS peak as well as the
flat band above the Fermi 
level as in carbides. Due to the shifting down of the Fermi level, the 
bands start to cross the M--K direction at $n$ = 2, and steadily increasing
numbers of bands 
cross the $\Gamma$--M and K--$\Gamma$ directions as $n$ increases.
Therefore, we observe a continuous increasing of \ce{N(E_{F})} in nitrides. We should
point out, for $n$ = 5, and 7, the Fermi level is located in the bottom of a
small valley, so the \ce{N(E_{F})} decreases a little compared to other nitrides.
Moreover, the DOS of carbides and nitrides show some characters 
as of bulk TiC and TiN when $n$ $\geqslant$ 7. However, the contribution
of C $2p$ orbital near the Fermi level is more significant in the thicker
MXenes than bulk. Thus, the thicker MXenes can still be considered as
2D materials.

\section{Conclusion}
In summary, we have systematically studied the structural and electronic properties
of functionalized \ce{Ti_{n+1}C_{n}} and \ce{Ti_{n+1}N_{n}} with $n$ up
to 9, using PBE and HSE06 functionals. We show that both PBE and HSE06
predict very similar structures and electronic structures. The HSE06 
functional predicts structural parameters
smaller than PBE by 0.02 $\sim$ 0.03 \AA and gives similar band structures. 
The functional groups are more likely to bond with
Ti atoms as the electron donor. Without terminations, MXenes are
magnetic, with the
magnetism primarily due to surface Ti atoms. Upon
functionalization, the magnetism is removed. All the functionalized 
MXenes are metallic except for semiconducting \ce{Ti_{2}C_{}O_{2}},
due to the strong Ti $3d$ and O $2p$ orbital hybridization near the 
Fermi level. The electronic properties of thicker MXenes ($n$ $\geqslant$ 5) 
are different from the thinnest, where the \ce{N(E_{F})} of thicker MXenes
is about 3.5 $\sim$ 8 times higher. This indicates that the 
thinnest and more thick MXenes are likely to display differing
properties. Our simulations shed some light
on selecting proper MXene systems for possible applications
and future experimental and theoretical studies are encouraged in
this field. 

\begin{acknowledgments}
  We thank Prof. Yury Gogotsi for helpful discussions and for bringing
  MXenes to our attention. V. Mochalin made helpful comments on the
  manuscript. This work was supported as part of the Fluid Interface
  Reactions, Structures and Transport (FIRST) Center, an Energy
  Frontier Research Center funded by the U.S. Department of Energy,
  Office of Science, Office of Basic Energy Sciences. This research
  used resources of the National Energy Research Scientific Computing
  Center, which is supported by the Office of Science of the
  U.S. Department of Energy under Contract No. DE-AC02-05CH11231.
\end{acknowledgments}

\bibliography{MXenes.bib}

\begin{thebibliography}{41}%
\makeatletter
\providecommand \@ifxundefined [1]{%
 \@ifx{#1\undefined}
}%
\providecommand \@ifnum [1]{%
 \ifnum #1\expandafter \@firstoftwo
 \else \expandafter \@secondoftwo
 \fi
}%
\providecommand \@ifx [1]{%
 \ifx #1\expandafter \@firstoftwo
 \else \expandafter \@secondoftwo
 \fi
}%
\providecommand \natexlab [1]{#1}%
\providecommand \enquote  [1]{``#1''}%
\providecommand \bibnamefont  [1]{#1}%
\providecommand \bibfnamefont [1]{#1}%
\providecommand \citenamefont [1]{#1}%
\providecommand \href@noop [0]{\@secondoftwo}%
\providecommand \href [0]{\begingroup \@sanitize@url \@href}%
\providecommand \@href[1]{\@@startlink{#1}\@@href}%
\providecommand \@@href[1]{\endgroup#1\@@endlink}%
\providecommand \@sanitize@url [0]{\catcode `\\12\catcode `\$12\catcode
  `\&12\catcode `\#12\catcode `\^12\catcode `\_12\catcode `\%12\relax}%
\providecommand \@@startlink[1]{}%
\providecommand \@@endlink[0]{}%
\providecommand \url  [0]{\begingroup\@sanitize@url \@url }%
\providecommand \@url [1]{\endgroup\@href {#1}{\urlprefix }}%
\providecommand \urlprefix  [0]{URL }%
\providecommand \Eprint [0]{\href }%
\providecommand \doibase [0]{http://dx.doi.org/}%
\providecommand \selectlanguage [0]{\@gobble}%
\providecommand \bibinfo  [0]{\@secondoftwo}%
\providecommand \bibfield  [0]{\@secondoftwo}%
\providecommand \translation [1]{[#1]}%
\providecommand \BibitemOpen [0]{}%
\providecommand \bibitemStop [0]{}%
\providecommand \bibitemNoStop [0]{.\EOS\space}%
\providecommand \EOS [0]{\spacefactor3000\relax}%
\providecommand \BibitemShut  [1]{\csname bibitem#1\endcsname}%
\let\auto@bib@innerbib\@empty
\bibitem [{\citenamefont {Novoselov}\ \emph {et~al.}(2004)\citenamefont
  {Novoselov}, \citenamefont {Geim}, \citenamefont {Morozov}, \citenamefont
  {Jiang}, \citenamefont {Zhang}, \citenamefont {Dubonos}, \citenamefont
  {Grigorieva},\ and\ \citenamefont {Firsov}}]{novoselov2004electric}%
  \BibitemOpen
  \bibfield  {author} {\bibinfo {author} {\bibfnamefont {K.}~\bibnamefont
  {Novoselov}}, \bibinfo {author} {\bibfnamefont {A.~K.}\ \bibnamefont {Geim}},
  \bibinfo {author} {\bibfnamefont {S.}~\bibnamefont {Morozov}}, \bibinfo
  {author} {\bibfnamefont {D.}~\bibnamefont {Jiang}}, \bibinfo {author}
  {\bibfnamefont {Y.}~\bibnamefont {Zhang}}, \bibinfo {author} {\bibfnamefont
  {S.}~\bibnamefont {Dubonos}}, \bibinfo {author} {\bibfnamefont
  {I.}~\bibnamefont {Grigorieva}}, \ and\ \bibinfo {author} {\bibfnamefont
  {A.}~\bibnamefont {Firsov}},\ }\href@noop {} {\bibfield  {journal} {\bibinfo
  {journal} {Science}\ }\textbf {\bibinfo {volume} {306}},\ \bibinfo {pages}
  {666} (\bibinfo {year} {2004})}\BibitemShut {NoStop}%
\bibitem [{\citenamefont {Berger}\ \emph {et~al.}(2006)\citenamefont {Berger},
  \citenamefont {Song}, \citenamefont {Li}, \citenamefont {Wu}, \citenamefont
  {Brown}, \citenamefont {Naud}, \citenamefont {Mayou}, \citenamefont {Li},
  \citenamefont {Hass}, \citenamefont {Marchenkov} \emph
  {et~al.}}]{berger2006electronic}%
  \BibitemOpen
  \bibfield  {author} {\bibinfo {author} {\bibfnamefont {C.}~\bibnamefont
  {Berger}}, \bibinfo {author} {\bibfnamefont {Z.}~\bibnamefont {Song}},
  \bibinfo {author} {\bibfnamefont {X.}~\bibnamefont {Li}}, \bibinfo {author}
  {\bibfnamefont {X.}~\bibnamefont {Wu}}, \bibinfo {author} {\bibfnamefont
  {N.}~\bibnamefont {Brown}}, \bibinfo {author} {\bibfnamefont
  {C.}~\bibnamefont {Naud}}, \bibinfo {author} {\bibfnamefont {D.}~\bibnamefont
  {Mayou}}, \bibinfo {author} {\bibfnamefont {T.}~\bibnamefont {Li}}, \bibinfo
  {author} {\bibfnamefont {J.}~\bibnamefont {Hass}}, \bibinfo {author}
  {\bibfnamefont {A.~N.}\ \bibnamefont {Marchenkov}},  \emph {et~al.},\
  }\href@noop {} {\bibfield  {journal} {\bibinfo  {journal} {Science}\ }\textbf
  {\bibinfo {volume} {312}},\ \bibinfo {pages} {1191} (\bibinfo {year}
  {2006})}\BibitemShut {NoStop}%
\bibitem [{\citenamefont {Zhang}\ \emph {et~al.}(2005)\citenamefont {Zhang},
  \citenamefont {Tan}, \citenamefont {Stormer},\ and\ \citenamefont
  {Kim}}]{zhang2005experimental}%
  \BibitemOpen
  \bibfield  {author} {\bibinfo {author} {\bibfnamefont {Y.}~\bibnamefont
  {Zhang}}, \bibinfo {author} {\bibfnamefont {Y.-W.}\ \bibnamefont {Tan}},
  \bibinfo {author} {\bibfnamefont {H.~L.}\ \bibnamefont {Stormer}}, \ and\
  \bibinfo {author} {\bibfnamefont {P.}~\bibnamefont {Kim}},\ }\href@noop {}
  {\bibfield  {journal} {\bibinfo  {journal} {Nature}\ }\textbf {\bibinfo
  {volume} {438}},\ \bibinfo {pages} {201} (\bibinfo {year}
  {2005})}\BibitemShut {NoStop}%
\bibitem [{\citenamefont {Stoller}\ \emph {et~al.}(2008)\citenamefont
  {Stoller}, \citenamefont {Park}, \citenamefont {Zhu}, \citenamefont {An},\
  and\ \citenamefont {Ruoff}}]{stoller2008graphene}%
  \BibitemOpen
  \bibfield  {author} {\bibinfo {author} {\bibfnamefont {M.~D.}\ \bibnamefont
  {Stoller}}, \bibinfo {author} {\bibfnamefont {S.}~\bibnamefont {Park}},
  \bibinfo {author} {\bibfnamefont {Y.}~\bibnamefont {Zhu}}, \bibinfo {author}
  {\bibfnamefont {J.}~\bibnamefont {An}}, \ and\ \bibinfo {author}
  {\bibfnamefont {R.~S.}\ \bibnamefont {Ruoff}},\ }\href@noop {} {\bibfield
  {journal} {\bibinfo  {journal} {Nano Lett.}\ }\textbf {\bibinfo {volume}
  {8}},\ \bibinfo {pages} {3498} (\bibinfo {year} {2008})}\BibitemShut
  {NoStop}%
\bibitem [{\citenamefont {Yoon}\ \emph {et~al.}(2011)\citenamefont {Yoon},
  \citenamefont {Ganapathi},\ and\ \citenamefont {Salahuddin}}]{yoon2011good}%
  \BibitemOpen
  \bibfield  {author} {\bibinfo {author} {\bibfnamefont {Y.}~\bibnamefont
  {Yoon}}, \bibinfo {author} {\bibfnamefont {K.}~\bibnamefont {Ganapathi}}, \
  and\ \bibinfo {author} {\bibfnamefont {S.}~\bibnamefont {Salahuddin}},\
  }\href@noop {} {\bibfield  {journal} {\bibinfo  {journal} {Nano Lett.}\
  }\textbf {\bibinfo {volume} {11}},\ \bibinfo {pages} {3768} (\bibinfo {year}
  {2011})}\BibitemShut {NoStop}%
\bibitem [{\citenamefont {Neto}\ \emph {et~al.}(2009)\citenamefont {Neto},
  \citenamefont {Guinea}, \citenamefont {Peres}, \citenamefont {Novoselov},\
  and\ \citenamefont {Geim}}]{neto2009electronic}%
  \BibitemOpen
  \bibfield  {author} {\bibinfo {author} {\bibfnamefont {A.~C.}\ \bibnamefont
  {Neto}}, \bibinfo {author} {\bibfnamefont {F.}~\bibnamefont {Guinea}},
  \bibinfo {author} {\bibfnamefont {N.}~\bibnamefont {Peres}}, \bibinfo
  {author} {\bibfnamefont {K.}~\bibnamefont {Novoselov}}, \ and\ \bibinfo
  {author} {\bibfnamefont {A.}~\bibnamefont {Geim}},\ }\href@noop {} {\bibfield
   {journal} {\bibinfo  {journal} {Rev. Mod. Phys.}\ }\textbf {\bibinfo
  {volume} {81}},\ \bibinfo {pages} {109} (\bibinfo {year} {2009})}\BibitemShut
  {NoStop}%
\bibitem [{\citenamefont {Yoo}\ \emph {et~al.}(2011)\citenamefont {Yoo},
  \citenamefont {Balakrishnan}, \citenamefont {Huang}, \citenamefont {Meunier},
  \citenamefont {Sumpter}, \citenamefont {Srivastava}, \citenamefont {Conway},
  \citenamefont {Mohana~Reddy}, \citenamefont {Yu}, \citenamefont {Vajtai}
  \emph {et~al.}}]{yoo2011ultrathin}%
  \BibitemOpen
  \bibfield  {author} {\bibinfo {author} {\bibfnamefont {J.~J.}\ \bibnamefont
  {Yoo}}, \bibinfo {author} {\bibfnamefont {K.}~\bibnamefont {Balakrishnan}},
  \bibinfo {author} {\bibfnamefont {J.}~\bibnamefont {Huang}}, \bibinfo
  {author} {\bibfnamefont {V.}~\bibnamefont {Meunier}}, \bibinfo {author}
  {\bibfnamefont {B.~G.}\ \bibnamefont {Sumpter}}, \bibinfo {author}
  {\bibfnamefont {A.}~\bibnamefont {Srivastava}}, \bibinfo {author}
  {\bibfnamefont {M.}~\bibnamefont {Conway}}, \bibinfo {author} {\bibfnamefont
  {A.~L.}\ \bibnamefont {Mohana~Reddy}}, \bibinfo {author} {\bibfnamefont
  {J.}~\bibnamefont {Yu}}, \bibinfo {author} {\bibfnamefont {R.}~\bibnamefont
  {Vajtai}},  \emph {et~al.},\ }\href@noop {} {\bibfield  {journal} {\bibinfo
  {journal} {Nano Lett.}\ }\textbf {\bibinfo {volume} {11}},\ \bibinfo {pages}
  {1423} (\bibinfo {year} {2011})}\BibitemShut {NoStop}%
\bibitem [{\citenamefont {Ramakrishna~Matte}\ \emph {et~al.}(2010)\citenamefont
  {Ramakrishna~Matte}, \citenamefont {Gomathi}, \citenamefont {Manna},
  \citenamefont {Late}, \citenamefont {Datta}, \citenamefont {Pati},\ and\
  \citenamefont {Rao}}]{ramakrishna2010mos2}%
  \BibitemOpen
  \bibfield  {author} {\bibinfo {author} {\bibfnamefont {H.}~\bibnamefont
  {Ramakrishna~Matte}}, \bibinfo {author} {\bibfnamefont {A.}~\bibnamefont
  {Gomathi}}, \bibinfo {author} {\bibfnamefont {A.~K.}\ \bibnamefont {Manna}},
  \bibinfo {author} {\bibfnamefont {D.~J.}\ \bibnamefont {Late}}, \bibinfo
  {author} {\bibfnamefont {R.}~\bibnamefont {Datta}}, \bibinfo {author}
  {\bibfnamefont {S.~K.}\ \bibnamefont {Pati}}, \ and\ \bibinfo {author}
  {\bibfnamefont {C.}~\bibnamefont {Rao}},\ }\href@noop {} {\bibfield
  {journal} {\bibinfo  {journal} {Angew. Chem. Int. Ed.}\ }\textbf {\bibinfo
  {volume} {122}},\ \bibinfo {pages} {4153} (\bibinfo {year}
  {2010})}\BibitemShut {NoStop}%
\bibitem [{\citenamefont {Pacile}\ \emph {et~al.}(2008)\citenamefont {Pacile},
  \citenamefont {Meyer}, \citenamefont {Girit},\ and\ \citenamefont
  {Zettl}}]{pacile2008two}%
  \BibitemOpen
  \bibfield  {author} {\bibinfo {author} {\bibfnamefont {D.}~\bibnamefont
  {Pacile}}, \bibinfo {author} {\bibfnamefont {J.}~\bibnamefont {Meyer}},
  \bibinfo {author} {\bibfnamefont {C.~O.}\ \bibnamefont {Girit}}, \ and\
  \bibinfo {author} {\bibfnamefont {A.}~\bibnamefont {Zettl}},\ }\href@noop {}
  {\bibfield  {journal} {\bibinfo  {journal} {Appl. Phys. Lett.}\ }\textbf
  {\bibinfo {volume} {92}},\ \bibinfo {pages} {133107} (\bibinfo {year}
  {2008})}\BibitemShut {NoStop}%
\bibitem [{\citenamefont {Novoselov}\ \emph {et~al.}(2005)\citenamefont
  {Novoselov}, \citenamefont {Jiang}, \citenamefont {Schedin}, \citenamefont
  {Booth}, \citenamefont {Khotkevich}, \citenamefont {Morozov},\ and\
  \citenamefont {Geim}}]{novoselov2005two}%
  \BibitemOpen
  \bibfield  {author} {\bibinfo {author} {\bibfnamefont {K.}~\bibnamefont
  {Novoselov}}, \bibinfo {author} {\bibfnamefont {D.}~\bibnamefont {Jiang}},
  \bibinfo {author} {\bibfnamefont {F.}~\bibnamefont {Schedin}}, \bibinfo
  {author} {\bibfnamefont {T.}~\bibnamefont {Booth}}, \bibinfo {author}
  {\bibfnamefont {V.}~\bibnamefont {Khotkevich}}, \bibinfo {author}
  {\bibfnamefont {S.}~\bibnamefont {Morozov}}, \ and\ \bibinfo {author}
  {\bibfnamefont {A.}~\bibnamefont {Geim}},\ }\href@noop {} {\bibfield
  {journal} {\bibinfo  {journal} {Proc. Natl. Acad. Sci. U.S.A.}\ }\textbf
  {\bibinfo {volume} {102}},\ \bibinfo {pages} {10451} (\bibinfo {year}
  {2005})}\BibitemShut {NoStop}%
\bibitem [{\citenamefont {Eklund}\ \emph {et~al.}(2010)\citenamefont {Eklund},
  \citenamefont {Beckers}, \citenamefont {Jansson}, \citenamefont
  {H{\"o}gberg},\ and\ \citenamefont {Hultman}}]{eklund2010m}%
  \BibitemOpen
  \bibfield  {author} {\bibinfo {author} {\bibfnamefont {P.}~\bibnamefont
  {Eklund}}, \bibinfo {author} {\bibfnamefont {M.}~\bibnamefont {Beckers}},
  \bibinfo {author} {\bibfnamefont {U.}~\bibnamefont {Jansson}}, \bibinfo
  {author} {\bibfnamefont {H.}~\bibnamefont {H{\"o}gberg}}, \ and\ \bibinfo
  {author} {\bibfnamefont {L.}~\bibnamefont {Hultman}},\ }\href@noop {}
  {\bibfield  {journal} {\bibinfo  {journal} {Thin Solid Films}\ }\textbf
  {\bibinfo {volume} {518}},\ \bibinfo {pages} {1851} (\bibinfo {year}
  {2010})}\BibitemShut {NoStop}%
\bibitem [{\citenamefont {Barsoum}(2000)}]{barsoum2000mn+}%
  \BibitemOpen
  \bibfield  {author} {\bibinfo {author} {\bibfnamefont {M.~W.}\ \bibnamefont
  {Barsoum}},\ }\href@noop {} {\bibfield  {journal} {\bibinfo  {journal} {Prog.
  Solid State Chem.}\ }\textbf {\bibinfo {volume} {28}},\ \bibinfo {pages}
  {201} (\bibinfo {year} {2000})}\BibitemShut {NoStop}%
\bibitem [{\citenamefont {Barsoum}\ and\ \citenamefont
  {Radovic}(2011)}]{barsoum2011elastic}%
  \BibitemOpen
  \bibfield  {author} {\bibinfo {author} {\bibfnamefont {M.~W.}\ \bibnamefont
  {Barsoum}}\ and\ \bibinfo {author} {\bibfnamefont {M.}~\bibnamefont
  {Radovic}},\ }\href@noop {} {\bibfield  {journal} {\bibinfo  {journal} {Annu.
  Rev. Mater. Sci.}\ }\textbf {\bibinfo {volume} {41}},\ \bibinfo {pages} {195}
  (\bibinfo {year} {2011})}\BibitemShut {NoStop}%
\bibitem [{\citenamefont {Wang}\ and\ \citenamefont
  {Zhou}(2010)}]{wang2010layered}%
  \BibitemOpen
  \bibfield  {author} {\bibinfo {author} {\bibfnamefont {X.}~\bibnamefont
  {Wang}}\ and\ \bibinfo {author} {\bibfnamefont {Y.}~\bibnamefont {Zhou}},\
  }\href@noop {} {\bibfield  {journal} {\bibinfo  {journal} {J. Mater. Sci.
  Technol.}\ }\textbf {\bibinfo {volume} {26}},\ \bibinfo {pages} {385}
  (\bibinfo {year} {2010})}\BibitemShut {NoStop}%
\bibitem [{\citenamefont {Sun}(2011)}]{sun2011progress}%
  \BibitemOpen
  \bibfield  {author} {\bibinfo {author} {\bibfnamefont {Z.}~\bibnamefont
  {Sun}},\ }\href@noop {} {\bibfield  {journal} {\bibinfo  {journal} {Int.
  Mater. Rev.}\ }\textbf {\bibinfo {volume} {56}},\ \bibinfo {pages} {143}
  (\bibinfo {year} {2011})}\BibitemShut {NoStop}%
\bibitem [{\citenamefont {Naguib}\ \emph {et~al.}(2011)\citenamefont {Naguib},
  \citenamefont {Kurtoglu}, \citenamefont {Presser}, \citenamefont {Lu},
  \citenamefont {Niu}, \citenamefont {Heon}, \citenamefont {Hultman},
  \citenamefont {Gogotsi},\ and\ \citenamefont {Barsoum}}]{naguib2011two}%
  \BibitemOpen
  \bibfield  {author} {\bibinfo {author} {\bibfnamefont {M.}~\bibnamefont
  {Naguib}}, \bibinfo {author} {\bibfnamefont {M.}~\bibnamefont {Kurtoglu}},
  \bibinfo {author} {\bibfnamefont {V.}~\bibnamefont {Presser}}, \bibinfo
  {author} {\bibfnamefont {J.}~\bibnamefont {Lu}}, \bibinfo {author}
  {\bibfnamefont {J.}~\bibnamefont {Niu}}, \bibinfo {author} {\bibfnamefont
  {M.}~\bibnamefont {Heon}}, \bibinfo {author} {\bibfnamefont {L.}~\bibnamefont
  {Hultman}}, \bibinfo {author} {\bibfnamefont {Y.}~\bibnamefont {Gogotsi}}, \
  and\ \bibinfo {author} {\bibfnamefont {M.~W.}\ \bibnamefont {Barsoum}},\
  }\href@noop {} {\bibfield  {journal} {\bibinfo  {journal} {Adv. Mater.}\
  }\textbf {\bibinfo {volume} {23}},\ \bibinfo {pages} {4248} (\bibinfo {year}
  {2011})}\BibitemShut {NoStop}%
\bibitem [{\citenamefont {Naguib}\ \emph
  {et~al.}(2012{\natexlab{a}})\citenamefont {Naguib}, \citenamefont
  {Mashtalir}, \citenamefont {Carle}, \citenamefont {Presser}, \citenamefont
  {Lu}, \citenamefont {Hultman}, \citenamefont {Gogotsi},\ and\ \citenamefont
  {Barsoum}}]{naguib2012two12}%
  \BibitemOpen
  \bibfield  {author} {\bibinfo {author} {\bibfnamefont {M.}~\bibnamefont
  {Naguib}}, \bibinfo {author} {\bibfnamefont {O.}~\bibnamefont {Mashtalir}},
  \bibinfo {author} {\bibfnamefont {J.}~\bibnamefont {Carle}}, \bibinfo
  {author} {\bibfnamefont {V.}~\bibnamefont {Presser}}, \bibinfo {author}
  {\bibfnamefont {J.}~\bibnamefont {Lu}}, \bibinfo {author} {\bibfnamefont
  {L.}~\bibnamefont {Hultman}}, \bibinfo {author} {\bibfnamefont
  {Y.}~\bibnamefont {Gogotsi}}, \ and\ \bibinfo {author} {\bibfnamefont
  {M.~W.}\ \bibnamefont {Barsoum}},\ }\href@noop {} {\bibfield  {journal}
  {\bibinfo  {journal} {ACS nano}\ }\textbf {\bibinfo {volume} {6}},\ \bibinfo
  {pages} {1322} (\bibinfo {year} {2012}{\natexlab{a}})}\BibitemShut {NoStop}%
\bibitem [{\citenamefont {Enyashin}\ and\ \citenamefont
  {Ivanovskii}(2012)}]{Enyashin201227}%
  \BibitemOpen
  \bibfield  {author} {\bibinfo {author} {\bibfnamefont {A.}~\bibnamefont
  {Enyashin}}\ and\ \bibinfo {author} {\bibfnamefont {A.}~\bibnamefont
  {Ivanovskii}},\ }\href {\doibase 10.1016/j.comptc.2012.02.034} {\bibfield
  {journal} {\bibinfo  {journal} {Comput. Theor. Chem.}\ }\textbf {\bibinfo
  {volume} {989}},\ \bibinfo {pages} {27 } (\bibinfo {year}
  {2012})}\BibitemShut {NoStop}%
\bibitem [{\citenamefont {Kurtoglu}\ \emph {et~al.}(2012)\citenamefont
  {Kurtoglu}, \citenamefont {Naguib}, \citenamefont {Gogotsi},\ and\
  \citenamefont {Barsoum}}]{kurtoglu2012first}%
  \BibitemOpen
  \bibfield  {author} {\bibinfo {author} {\bibfnamefont {M.}~\bibnamefont
  {Kurtoglu}}, \bibinfo {author} {\bibfnamefont {M.}~\bibnamefont {Naguib}},
  \bibinfo {author} {\bibfnamefont {Y.}~\bibnamefont {Gogotsi}}, \ and\
  \bibinfo {author} {\bibfnamefont {M.~W.}\ \bibnamefont {Barsoum}},\
  }\href@noop {} {\bibfield  {journal} {\bibinfo  {journal} {MRS Commun.}\
  }\textbf {\bibinfo {volume} {2}},\ \bibinfo {pages} {133} (\bibinfo {year}
  {2012})}\BibitemShut {NoStop}%
\bibitem [{\citenamefont {Come}\ \emph {et~al.}(2012)\citenamefont {Come},
  \citenamefont {Naguib}, \citenamefont {Rozier}, \citenamefont {Barsoum},
  \citenamefont {Gogotsi}, \citenamefont {Taberna}, \citenamefont {Morcrette},\
  and\ \citenamefont {Simon}}]{come2012non}%
  \BibitemOpen
  \bibfield  {author} {\bibinfo {author} {\bibfnamefont {J.}~\bibnamefont
  {Come}}, \bibinfo {author} {\bibfnamefont {M.}~\bibnamefont {Naguib}},
  \bibinfo {author} {\bibfnamefont {P.}~\bibnamefont {Rozier}}, \bibinfo
  {author} {\bibfnamefont {M.}~\bibnamefont {Barsoum}}, \bibinfo {author}
  {\bibfnamefont {Y.}~\bibnamefont {Gogotsi}}, \bibinfo {author} {\bibfnamefont
  {P.-L.}\ \bibnamefont {Taberna}}, \bibinfo {author} {\bibfnamefont
  {M.}~\bibnamefont {Morcrette}}, \ and\ \bibinfo {author} {\bibfnamefont
  {P.}~\bibnamefont {Simon}},\ }\href@noop {} {\bibfield  {journal} {\bibinfo
  {journal} {J. Electrochem. Soc.}\ }\textbf {\bibinfo {volume} {159}},\
  \bibinfo {pages} {A1368} (\bibinfo {year} {2012})}\BibitemShut {NoStop}%
\bibitem [{\citenamefont {Naguib}\ \emph
  {et~al.}(2012{\natexlab{b}})\citenamefont {Naguib}, \citenamefont {Come},
  \citenamefont {Dyatkin}, \citenamefont {Presser}, \citenamefont {Taberna},
  \citenamefont {Simon}, \citenamefont {Barsoum},\ and\ \citenamefont
  {Gogotsi}}]{naguib2012mxene}%
  \BibitemOpen
  \bibfield  {author} {\bibinfo {author} {\bibfnamefont {M.}~\bibnamefont
  {Naguib}}, \bibinfo {author} {\bibfnamefont {J.}~\bibnamefont {Come}},
  \bibinfo {author} {\bibfnamefont {B.}~\bibnamefont {Dyatkin}}, \bibinfo
  {author} {\bibfnamefont {V.}~\bibnamefont {Presser}}, \bibinfo {author}
  {\bibfnamefont {P.-L.}\ \bibnamefont {Taberna}}, \bibinfo {author}
  {\bibfnamefont {P.}~\bibnamefont {Simon}}, \bibinfo {author} {\bibfnamefont
  {M.~W.}\ \bibnamefont {Barsoum}}, \ and\ \bibinfo {author} {\bibfnamefont
  {Y.}~\bibnamefont {Gogotsi}},\ }\href@noop {} {\bibfield  {journal} {\bibinfo
   {journal} {Electrochem. Commun.}\ }\textbf {\bibinfo {volume} {16}},\
  \bibinfo {pages} {61} (\bibinfo {year} {2012}{\natexlab{b}})}\BibitemShut
  {NoStop}%
\bibitem [{\citenamefont {Tang}\ \emph {et~al.}(2012)\citenamefont {Tang},
  \citenamefont {Zhou},\ and\ \citenamefont {Shen}}]{tang2012mxenes}%
  \BibitemOpen
  \bibfield  {author} {\bibinfo {author} {\bibfnamefont {Q.}~\bibnamefont
  {Tang}}, \bibinfo {author} {\bibfnamefont {Z.}~\bibnamefont {Zhou}}, \ and\
  \bibinfo {author} {\bibfnamefont {P.}~\bibnamefont {Shen}},\ }\href@noop {}
  {\bibfield  {journal} {\bibinfo  {journal} {J. Am. Chem. Soc.}\ }\textbf
  {\bibinfo {volume} {134}},\ \bibinfo {pages} {16909} (\bibinfo {year}
  {2012})}\BibitemShut {NoStop}%
\bibitem [{\citenamefont {Mashtalir}\ \emph {et~al.}(2013)\citenamefont
  {Mashtalir}, \citenamefont {Naguib}, \citenamefont {Mochalin}, \citenamefont
  {DallAgnese}, \citenamefont {Heon}, \citenamefont {Barsoum},\ and\
  \citenamefont {Gogotsi}}]{mashtalir2013inter}%
  \BibitemOpen
  \bibfield  {author} {\bibinfo {author} {\bibfnamefont {O.}~\bibnamefont
  {Mashtalir}}, \bibinfo {author} {\bibfnamefont {M.}~\bibnamefont {Naguib}},
  \bibinfo {author} {\bibfnamefont {V.~N.}\ \bibnamefont {Mochalin}}, \bibinfo
  {author} {\bibfnamefont {Y.}~\bibnamefont {DallAgnese}}, \bibinfo {author}
  {\bibfnamefont {M.}~\bibnamefont {Heon}}, \bibinfo {author} {\bibfnamefont
  {M.~W.}\ \bibnamefont {Barsoum}}, \ and\ \bibinfo {author} {\bibfnamefont
  {Y.}~\bibnamefont {Gogotsi}},\ }\href@noop {} {\bibfield  {journal} {\bibinfo
   {journal} {Nat. Commun.}\ }\textbf {\bibinfo {volume} {4}},\ \bibinfo
  {pages} {1716} (\bibinfo {year} {2013})}\BibitemShut {NoStop}%
\bibitem [{\citenamefont {Khazaei}\ \emph {et~al.}(2013)\citenamefont
  {Khazaei}, \citenamefont {Arai}, \citenamefont {Sasaki}, \citenamefont
  {Chung}, \citenamefont {Venkataramanan}, \citenamefont {Estili},
  \citenamefont {Sakka},\ and\ \citenamefont {Kawazoe}}]{khazaei2012novel}%
  \BibitemOpen
  \bibfield  {author} {\bibinfo {author} {\bibfnamefont {M.}~\bibnamefont
  {Khazaei}}, \bibinfo {author} {\bibfnamefont {M.}~\bibnamefont {Arai}},
  \bibinfo {author} {\bibfnamefont {T.}~\bibnamefont {Sasaki}}, \bibinfo
  {author} {\bibfnamefont {C.-Y.}\ \bibnamefont {Chung}}, \bibinfo {author}
  {\bibfnamefont {N.~S.}\ \bibnamefont {Venkataramanan}}, \bibinfo {author}
  {\bibfnamefont {M.}~\bibnamefont {Estili}}, \bibinfo {author} {\bibfnamefont
  {Y.}~\bibnamefont {Sakka}}, \ and\ \bibinfo {author} {\bibfnamefont
  {Y.}~\bibnamefont {Kawazoe}},\ }\href {\doibase 10.1002/adfm.201202502}
  {\bibfield  {journal} {\bibinfo  {journal} {Adv. Funct. Mater.}\ }\textbf
  {\bibinfo {volume} {23}},\ \bibinfo {pages} {2185} (\bibinfo {year}
  {2013})}\BibitemShut {NoStop}%
\bibitem [{\citenamefont {Shein}\ and\ \citenamefont
  {Ivanovskii}(2012)}]{shein2012graphene}%
  \BibitemOpen
  \bibfield  {author} {\bibinfo {author} {\bibfnamefont {I.}~\bibnamefont
  {Shein}}\ and\ \bibinfo {author} {\bibfnamefont {A.}~\bibnamefont
  {Ivanovskii}},\ }\href@noop {} {\bibfield  {journal} {\bibinfo  {journal}
  {Comput. Mater. Sci.}\ }\textbf {\bibinfo {volume} {65}},\ \bibinfo {pages}
  {104} (\bibinfo {year} {2012})}\BibitemShut {NoStop}%
\bibitem [{\citenamefont {Lane}\ \emph {et~al.}(2013)\citenamefont {Lane},
  \citenamefont {Barsoum},\ and\ \citenamefont
  {Rondinelli}}]{lane2013correlation}%
  \BibitemOpen
  \bibfield  {author} {\bibinfo {author} {\bibfnamefont {N.~J.}\ \bibnamefont
  {Lane}}, \bibinfo {author} {\bibfnamefont {M.~W.}\ \bibnamefont {Barsoum}}, \
  and\ \bibinfo {author} {\bibfnamefont {J.~M.}\ \bibnamefont {Rondinelli}},\
  }\href@noop {} {\bibfield  {journal} {\bibinfo  {journal} {EPL}\ }\textbf
  {\bibinfo {volume} {101}},\ \bibinfo {pages} {57004} (\bibinfo {year}
  {2013})}\BibitemShut {NoStop}%
\bibitem [{\citenamefont {Kresse}\ and\ \citenamefont
  {Furthm{\"u}ller}(1996)}]{kresse1996efficient}%
  \BibitemOpen
  \bibfield  {author} {\bibinfo {author} {\bibfnamefont {G.}~\bibnamefont
  {Kresse}}\ and\ \bibinfo {author} {\bibfnamefont {J.}~\bibnamefont
  {Furthm{\"u}ller}},\ }\href@noop {} {\bibfield  {journal} {\bibinfo
  {journal} {Phys. Rev. B}\ }\textbf {\bibinfo {volume} {54}},\ \bibinfo
  {pages} {11169} (\bibinfo {year} {1996})}\BibitemShut {NoStop}%
\bibitem [{\citenamefont {Bl{\"o}chl}(1994)}]{blochl1994projector}%
  \BibitemOpen
  \bibfield  {author} {\bibinfo {author} {\bibfnamefont {P.~E.}\ \bibnamefont
  {Bl{\"o}chl}},\ }\href@noop {} {\bibfield  {journal} {\bibinfo  {journal}
  {Phys. Rev. B}\ }\textbf {\bibinfo {volume} {50}},\ \bibinfo {pages} {17953}
  (\bibinfo {year} {1994})}\BibitemShut {NoStop}%
\bibitem [{\citenamefont {Perdew}\ \emph {et~al.}(1996)\citenamefont {Perdew},
  \citenamefont {Burke},\ and\ \citenamefont
  {Ernzerhof}}]{perdew1996generalized}%
  \BibitemOpen
  \bibfield  {author} {\bibinfo {author} {\bibfnamefont {J.~P.}\ \bibnamefont
  {Perdew}}, \bibinfo {author} {\bibfnamefont {K.}~\bibnamefont {Burke}}, \
  and\ \bibinfo {author} {\bibfnamefont {M.}~\bibnamefont {Ernzerhof}},\
  }\href@noop {} {\bibfield  {journal} {\bibinfo  {journal} {Phys. Rev. Lett.}\
  }\textbf {\bibinfo {volume} {77}},\ \bibinfo {pages} {3865} (\bibinfo {year}
  {1996})}\BibitemShut {NoStop}%
\bibitem [{\citenamefont {Heyd}\ \emph {et~al.}(2003)\citenamefont {Heyd},
  \citenamefont {Scuseria},\ and\ \citenamefont {Ernzerhof}}]{heyd8207}%
  \BibitemOpen
  \bibfield  {author} {\bibinfo {author} {\bibfnamefont {J.}~\bibnamefont
  {Heyd}}, \bibinfo {author} {\bibfnamefont {G.~E.}\ \bibnamefont {Scuseria}},
  \ and\ \bibinfo {author} {\bibfnamefont {M.}~\bibnamefont {Ernzerhof}},\
  }\href {\doibase 10.1063/1.1564060} {\bibfield  {journal} {\bibinfo
  {journal} {J. Chem. Phys.}\ }\textbf {\bibinfo {volume} {118}},\ \bibinfo
  {pages} {8207} (\bibinfo {year} {2003})}\BibitemShut {NoStop}%
\bibitem [{\citenamefont {Paier}\ \emph {et~al.}(2006)\citenamefont {Paier},
  \citenamefont {Marsman}, \citenamefont {Hummer}, \citenamefont {Kresse},
  \citenamefont {Gerber},\ and\ \citenamefont
  {{\'A}ngy{\'a}n}}]{paier2006screened}%
  \BibitemOpen
  \bibfield  {author} {\bibinfo {author} {\bibfnamefont {J.}~\bibnamefont
  {Paier}}, \bibinfo {author} {\bibfnamefont {M.}~\bibnamefont {Marsman}},
  \bibinfo {author} {\bibfnamefont {K.}~\bibnamefont {Hummer}}, \bibinfo
  {author} {\bibfnamefont {G.}~\bibnamefont {Kresse}}, \bibinfo {author}
  {\bibfnamefont {I.~C.}\ \bibnamefont {Gerber}}, \ and\ \bibinfo {author}
  {\bibfnamefont {J.~G.}\ \bibnamefont {{\'A}ngy{\'a}n}},\ }\href@noop {}
  {\bibfield  {journal} {\bibinfo  {journal} {J. Chem. Phys.}\ }\textbf
  {\bibinfo {volume} {124}},\ \bibinfo {pages} {154709} (\bibinfo {year}
  {2006})}\BibitemShut {NoStop}%
\bibitem [{\citenamefont {Heyd}\ \emph {et~al.}(2006)\citenamefont {Heyd},
  \citenamefont {Scuseria},\ and\ \citenamefont {Ernzerhof}}]{heyd2006erratum}%
  \BibitemOpen
  \bibfield  {author} {\bibinfo {author} {\bibfnamefont {J.}~\bibnamefont
  {Heyd}}, \bibinfo {author} {\bibfnamefont {G.~E.}\ \bibnamefont {Scuseria}},
  \ and\ \bibinfo {author} {\bibfnamefont {M.}~\bibnamefont {Ernzerhof}},\
  }\href@noop {} {\bibfield  {journal} {\bibinfo  {journal} {J. Chem. Phys.}\
  }\textbf {\bibinfo {volume} {124}},\ \bibinfo {pages} {219906} (\bibinfo
  {year} {2006})}\BibitemShut {NoStop}%
\bibitem [{\citenamefont {Eyert}(2011)}]{eyert2011vo}%
  \BibitemOpen
  \bibfield  {author} {\bibinfo {author} {\bibfnamefont {V.}~\bibnamefont
  {Eyert}},\ }\href@noop {} {\bibfield  {journal} {\bibinfo  {journal} {Phys.
  Rev. Lett.}\ }\textbf {\bibinfo {volume} {107}},\ \bibinfo {pages} {016401}
  (\bibinfo {year} {2011})}\BibitemShut {NoStop}%
\bibitem [{\citenamefont {Manoun}\ \emph {et~al.}(2006)\citenamefont {Manoun},
  \citenamefont {Zhang}, \citenamefont {Saxena}, \citenamefont {El-Raghy},\
  and\ \citenamefont {Barsoum}}]{manoun2006x}%
  \BibitemOpen
  \bibfield  {author} {\bibinfo {author} {\bibfnamefont {B.}~\bibnamefont
  {Manoun}}, \bibinfo {author} {\bibfnamefont {F.}~\bibnamefont {Zhang}},
  \bibinfo {author} {\bibfnamefont {S.}~\bibnamefont {Saxena}}, \bibinfo
  {author} {\bibfnamefont {T.}~\bibnamefont {El-Raghy}}, \ and\ \bibinfo
  {author} {\bibfnamefont {M.}~\bibnamefont {Barsoum}},\ }\href@noop {}
  {\bibfield  {journal} {\bibinfo  {journal} {J. Phys. Chem. Solids}\ }\textbf
  {\bibinfo {volume} {67}},\ \bibinfo {pages} {2091} (\bibinfo {year}
  {2006})}\BibitemShut {NoStop}%
\bibitem [{\citenamefont {H{\"o}gberg}\ \emph {et~al.}(2005)\citenamefont
  {H{\"o}gberg}, \citenamefont {Eklund}, \citenamefont {Emmerlich},
  \citenamefont {Birch},\ and\ \citenamefont {Hultman}}]{hogberg2005epitaxial}%
  \BibitemOpen
  \bibfield  {author} {\bibinfo {author} {\bibfnamefont {H.}~\bibnamefont
  {H{\"o}gberg}}, \bibinfo {author} {\bibfnamefont {P.}~\bibnamefont {Eklund}},
  \bibinfo {author} {\bibfnamefont {J.}~\bibnamefont {Emmerlich}}, \bibinfo
  {author} {\bibfnamefont {J.}~\bibnamefont {Birch}}, \ and\ \bibinfo {author}
  {\bibfnamefont {L.}~\bibnamefont {Hultman}},\ }\href@noop {} {\bibfield
  {journal} {\bibinfo  {journal} {J. Mater. Res.}\ }\textbf {\bibinfo {volume}
  {20}},\ \bibinfo {pages} {779} (\bibinfo {year} {2005})}\BibitemShut
  {NoStop}%
\bibitem [{\citenamefont {Zheng}\ \emph {et~al.}(2010)\citenamefont {Zheng},
  \citenamefont {Wang}, \citenamefont {Lu}, \citenamefont {Li}, \citenamefont
  {Wang},\ and\ \citenamefont {Zhou}}]{zheng2010ti0}%
  \BibitemOpen
  \bibfield  {author} {\bibinfo {author} {\bibfnamefont {L.}~\bibnamefont
  {Zheng}}, \bibinfo {author} {\bibfnamefont {J.}~\bibnamefont {Wang}},
  \bibinfo {author} {\bibfnamefont {X.}~\bibnamefont {Lu}}, \bibinfo {author}
  {\bibfnamefont {F.}~\bibnamefont {Li}}, \bibinfo {author} {\bibfnamefont
  {J.}~\bibnamefont {Wang}}, \ and\ \bibinfo {author} {\bibfnamefont
  {Y.}~\bibnamefont {Zhou}},\ }\href@noop {} {\bibfield  {journal} {\bibinfo
  {journal} {J. Am. Ceram. Soc.}\ }\textbf {\bibinfo {volume} {93}},\ \bibinfo
  {pages} {3068} (\bibinfo {year} {2010})}\BibitemShut {NoStop}%
\bibitem [{\citenamefont {Zhang}\ \emph {et~al.}(2009)\citenamefont {Zhang},
  \citenamefont {Liu}, \citenamefont {Wang},\ and\ \citenamefont
  {Zhou}}]{zhang2009low}%
  \BibitemOpen
  \bibfield  {author} {\bibinfo {author} {\bibfnamefont {J.}~\bibnamefont
  {Zhang}}, \bibinfo {author} {\bibfnamefont {B.}~\bibnamefont {Liu}}, \bibinfo
  {author} {\bibfnamefont {J.}~\bibnamefont {Wang}}, \ and\ \bibinfo {author}
  {\bibfnamefont {Y.}~\bibnamefont {Zhou}},\ }\href@noop {} {\bibfield
  {journal} {\bibinfo  {journal} {J. Mater. Res.}\ }\textbf {\bibinfo {volume}
  {24}},\ \bibinfo {pages} {39} (\bibinfo {year} {2009})}\BibitemShut {NoStop}%
\bibitem [{\citenamefont {Barsoum}\ \emph {et~al.}(2000)\citenamefont
  {Barsoum}, \citenamefont {Rawn}, \citenamefont {El-Raghy}, \citenamefont
  {Procopio}, \citenamefont {Porter}, \citenamefont {Wang},\ and\ \citenamefont
  {Hubbard}}]{barsoum2000thermal}%
  \BibitemOpen
  \bibfield  {author} {\bibinfo {author} {\bibfnamefont {M.}~\bibnamefont
  {Barsoum}}, \bibinfo {author} {\bibfnamefont {C.}~\bibnamefont {Rawn}},
  \bibinfo {author} {\bibfnamefont {T.}~\bibnamefont {El-Raghy}}, \bibinfo
  {author} {\bibfnamefont {A.}~\bibnamefont {Procopio}}, \bibinfo {author}
  {\bibfnamefont {W.}~\bibnamefont {Porter}}, \bibinfo {author} {\bibfnamefont
  {H.}~\bibnamefont {Wang}}, \ and\ \bibinfo {author} {\bibfnamefont
  {C.}~\bibnamefont {Hubbard}},\ }\href@noop {} {\bibfield  {journal} {\bibinfo
   {journal} {J. Appl. Phys.}\ }\textbf {\bibinfo {volume} {87}},\ \bibinfo
  {pages} {8407} (\bibinfo {year} {2000})}\BibitemShut {NoStop}%
\bibitem [{\citenamefont {Li}\ and\ \citenamefont {Wang}(2010)}]{li2010first}%
  \BibitemOpen
  \bibfield  {author} {\bibinfo {author} {\bibfnamefont {C.}~\bibnamefont
  {Li}}\ and\ \bibinfo {author} {\bibfnamefont {Z.}~\bibnamefont {Wang}},\
  }\href@noop {} {\bibfield  {journal} {\bibinfo  {journal} {J. Appl. Phys.}\
  }\textbf {\bibinfo {volume} {107}},\ \bibinfo {pages} {123511} (\bibinfo
  {year} {2010})}\BibitemShut {NoStop}%
\bibitem [{\citenamefont {Keast}\ \emph {et~al.}(2009)\citenamefont {Keast},
  \citenamefont {Harris},\ and\ \citenamefont {Smith}}]{keast2009prediction}%
  \BibitemOpen
  \bibfield  {author} {\bibinfo {author} {\bibfnamefont {V.}~\bibnamefont
  {Keast}}, \bibinfo {author} {\bibfnamefont {S.}~\bibnamefont {Harris}}, \
  and\ \bibinfo {author} {\bibfnamefont {D.}~\bibnamefont {Smith}},\
  }\href@noop {} {\bibfield  {journal} {\bibinfo  {journal} {Phys. Rev. B}\
  }\textbf {\bibinfo {volume} {80}},\ \bibinfo {pages} {214113} (\bibinfo
  {year} {2009})}\BibitemShut {NoStop}%
\bibitem [{\citenamefont {Hou}\ \emph {et~al.}(2011)\citenamefont {Hou},
  \citenamefont {Wang}, \citenamefont {Ikeda}, \citenamefont {Huang},
  \citenamefont {Terakura}, \citenamefont {Boero}, \citenamefont {Oshima},
  \citenamefont {Kakimoto},\ and\ \citenamefont {Miyata}}]{hou2011effect}%
  \BibitemOpen
  \bibfield  {author} {\bibinfo {author} {\bibfnamefont {Z.}~\bibnamefont
  {Hou}}, \bibinfo {author} {\bibfnamefont {X.}~\bibnamefont {Wang}}, \bibinfo
  {author} {\bibfnamefont {T.}~\bibnamefont {Ikeda}}, \bibinfo {author}
  {\bibfnamefont {S.-F.}\ \bibnamefont {Huang}}, \bibinfo {author}
  {\bibfnamefont {K.}~\bibnamefont {Terakura}}, \bibinfo {author}
  {\bibfnamefont {M.}~\bibnamefont {Boero}}, \bibinfo {author} {\bibfnamefont
  {M.}~\bibnamefont {Oshima}}, \bibinfo {author} {\bibfnamefont {M.-a.}\
  \bibnamefont {Kakimoto}}, \ and\ \bibinfo {author} {\bibfnamefont
  {S.}~\bibnamefont {Miyata}},\ }\href@noop {} {\bibfield  {journal} {\bibinfo
  {journal} {J. Phys. Chem. C}\ }\textbf {\bibinfo {volume} {115}},\ \bibinfo
  {pages} {5392} (\bibinfo {year} {2011})}\BibitemShut {NoStop}%
\end{thebibliography}%

\end{document}